\documentclass[a4paper,fleqn]{cas-sc}

\usepackage[square,numbers,compress,sort]{natbib}
\makeatletter
\renewcommand\@biblabel[1]{#1.}
\makeatother
\usepackage{csquotes}
\usepackage{xcolor}
\usepackage[colorinlistoftodos,prependcaption]{todonotes}
\usepackage[inline]{enumitem}
\usepackage{xspace}
\usepackage{multicol}
\usepackage{color}
\usepackage{xcolor}
\usepackage{acronym} 
\usepackage{subcaption}
\usepackage{longtable}
\usepackage{afterpage}
\usepackage{csquotes}

\usepackage{url}
\usepackage{amssymb}%
\usepackage{wasysym}
\usepackage{makecell}
\usepackage{pifont}

\usepackage[commandnameprefix=ifneeded]{changes}

\def\tsc#1{\csdef{#1}{\textsc{\lowercase{#1}}\xspace}}
\tsc{WGM}
\tsc{QE}
\tsc{EP}
\tsc{PMS}
\tsc{BEC}
\tsc{DE}

\usepackage{float}
\usepackage[usenames,dvipsnames]{pstricks}
\usepackage{pstricks-add}
 \usepackage{epsfig}
 \usepackage{tablefootnote}
 \usepackage{pst-grad} 
 \usepackage{pst-plot} 
 \usepackage[space]{grffile} 
 \usepackage{etoolbox} 
 \makeatletter 
 \patchcmd\Gread@eps{\@inputcheck#1 }{\@inputcheck"#1"\relax}{}{}
 \makeatother
\usepackage{verbatim}
\usepackage{multicol}
\usepackage[ruled,lined,noresetcount ]{algorithm2e}
\usepackage{algpseudocode}
\usepackage{mathtools}

\begin{document}
\let\WriteBookmarks\relax
\def\floatpagepagefraction{1}
\def\textpagefraction{.001}

\shorttitle{Transformers and Large Language Models for Efficient Intrusion Detection Systems: A Comprehensive Survey}

\shortauthors{Kheddar et~al.}
                      
\title [mode = title]{ Transformers and Large Language Models for Efficient Intrusion Detection Systems: A Comprehensive Survey }

\vskip2mm

\author[1]{Hamza Kheddar\corref{cor1}}
[orcid=0000-0002-9532-2453]
\ead{kheddar.hamza@univ-medea.dz}
\cormark[1]
\credit{Conceptualization; Methodology; Data Curation; Resources; Investigation; Project management; Visualization; Writing original draft; Writing; review and editing; Validation; Project administration; Supervision.}

\address[1]{LSEA Laboratory, Department of Electrical Engineering, University of Medea, 26000, Algeria}

\tnotetext[1]{The first is the corresponding author.}

\begin{abstract}
With significant advancements in Transformers and large language models (LLMs), natural language processing (NLP) has extended its reach into many research fields due to its enhanced capabilities in text generation and user interaction. One field benefiting greatly from these advancements is cybersecurity. In cybersecurity, many parameters that need to be protected and exchanged between senders and receivers are in the form of text and tabular data, making NLP a valuable tool in enhancing the security measures of communication protocols. This survey paper provides a comprehensive analysis of the utilization of Transformers and LLMs in cyber-threat detection systems. The methodology of paper selection and bibliometric analysis is outlined to establish a rigorous framework for evaluating existing research. The fundamentals of Transformers are discussed, including background information on various cyber-attacks and datasets commonly used in this field. The survey explores the application of Transformers in intrusion detection systems (IDSs), focusing on different architectures such as Attention-based models, LLMs like BERT and GPT, CNN/LSTM-Transformer hybrids, and emerging approaches like Vision Transformers (ViTs), and more. Furthermore, it explores the diverse environments and applications where Transformers and LLMs-based IDS have been implemented, including computer networks, internet-of-things (IoT) devices, critical infrastructure protection, cloud computing, software-defined networking (SDN), as well as in autonomous vehicles (AVs). The paper also addresses research challenges and future directions in this area, identifying key issues such as interpretability, scalability, and adaptability to evolving threats, and more. Finally, the conclusion summarizes the findings and highlights the significance of Transformers and LLMs in enhancing cyber-threat detection capabilities, while also outlining potential avenues for further research and development.
\end{abstract}




\maketitle


\section{Introduction} \label{sec1}


In today's swiftly evolving network ecosystem, characterized by the emergence of technologies like 5G and the widespread adoption of the \ac{IoT} \cite{kheddar2022efficient}, the potential for threats and vulnerabilities has expanded significantly. Consequently, concerns regarding network security are escalating. {Network security attacks are diverse and continuously evolving, posing significant threats to the integrity, confidentiality, and availability of information systems. These attacks can be broadly categorized into various types, including malware, phishing, \ac{DoS} attacks, \ac{MitM} attacks, and \ac{APTs} \cite{sharma2023advanced}. Malware encompasses viruses, worms, ransomware, and spyware that infiltrate systems to steal data, disrupt operations, or demand ransoms. Phishing attacks deceive users into revealing sensitive information through fraudulent emails or websites. DoS attacks overwhelm systems with excessive traffic, rendering services unavailable. \ac{MitM} attacks intercept and manipulate communications between parties, while \ac{APTs} involve prolonged and targeted cyber-espionage campaigns against specific organizations or individuals.
To combat these threats, various attack prevention methods have been developed, each with its unique approach and effectiveness. These methods include firewalls, antivirus software, encryption, \ac{IDS} \cite{kheddar2023deep}, \ac{IPS} \cite{kumar2022intrusion}, and \ac{SIEM} \cite{aguirre2011improving} systems. Firewalls act as a barrier between trusted and untrusted networks, controlling incoming and outgoing traffic based on predefined security rules. Antivirus software detects and removes malicious software, while encryption ensures data confidentiality by converting information into unreadable code without the correct decryption key. \ac{IDS} and \ac{IPS} monitor network traffic for suspicious activities, with \ac{IDS} alerting administrators of potential threats and \ac{IPS} actively blocking malicious traffic. \ac{SIEM} systems aggregate and analyze log data from various sources to detect and respond to security incidents in real-time. Among these, \ac{IDS} has advanced significantly with \ac{ML} and \ac{DL} integration, enhancing \ac{HIDS} and \ac{NIDS} \cite{kheddar2023deep}. \ac{HIDS} monitors individual devices, focusing on system logs and behaviors to detect threats like unauthorized file changes. However, it is limited to specific hosts. \ac{NIDS}, conversely, analyzes network traffic to identify threats like \ac{DDoS} attacks and network scanning, offering a broader security perspective. \ac{AI} enhances both categories by analyzing vast data to recognize complex threat patterns, improving detection accuracy and adaptability to evolving threats.}  Their versatility stems from the abundance of network data available for training intrusion scenarios and crafting \ac{AI}-driven  \ac{IDS} models. Furthermore, advancements in technology have reinforced computational capabilities, facilitating faster and more cost-effective model training. The widespread adoption of \ac{DL} ensures precise model optimization through continuous self-learning.

 

\begin{table}[!htbp]
\centering
{\small \section*{Acronyms and Abbreviations}}
\begin{multicols}{3}
\footnotesize
\begin{acronym}[BiAALSTM] 
\acro{Acc}{accuracy}
\acro{AI}{artificial intelligence}
\acro{APTs}{advanced persistent threats}
\acro{AS}{alert score}
\acro{AV}{autonomous vehicles}
\acro{BERT}{bidirectional encoder representations from transformers}
\acro{Bi-LSTM}{bi-directional long short-term memory}
\acro{CAN}{controller area network}
\acro{CANINE}{character architecture with no tokenization in neural encoders}
\acro{CGAN}{conditional generative adversarial network}
\acro{CIA}{compositional instruction attack}
\acro{CNN}{convolution neural network}
\acro{CTI}{cyber threat intelligence}
\acro{CV}{computer vision}
\acro{DDoS}{distributed denial-of-service}
\acro{DL}{deep learning}
\acro{DNN}{deep neural network}
\acro{DNS}{domain name system}
\acro{DoS}{denial-of-service}
\acro{DT}{decision tree}
\acro{F1}{F1-score}
\acro{FAR}{false alarm rate}
\acro{FFNN}{feedforward neural network}
\acro{FL}{federated learning}
\acro{FPR}{false positive rate}
\acro{FPGA}{field programmable gate array}
\acro{FR}{fooling rate}
\acro{FTP}{File Transfer Protocol}
\acro{GAN}{generative adversarial network}
\acro{GDPR}{general data protection regulation}
\acro{GMM}{Gaussian mixture model}
\acro{GPT}{generative pre-trained transformer}
\acro{GRU}{gated recurrent unit}
\acro{GSF}{genetically seeded flora}
\acro{HIDS}{host intrusion detection system}
\acro{HTTP}{HyperText Transfer Protocol}
\acro{ICMP}{Internet Control Message Protocol}
\acro{ICS}{industrial control system}
\acro{IDS}{intrusion detection system}
\acro{IG}{information gain}
\acro{IIoT}{industrial internet of things}
\acro{IMAP}{Internet message access protocol}
\acro{IoT}{Internet-of-things}
\acro{IoV}{Internet over vehicle}
\acro{IP}{Internet Protocol}
\acro{IPS}{intrusion prevention systems}
\acro{KLD}{Kullback-Leibler divergence}
\acro{LIME}{local interpretable model-agnostic explanations}
\acro{LLM}{large language model}
\acro{LSTM}{long short-term memory}
\acro{MAC}{media access control}
\acro{MCC}{Matthew’s correlation coefficient}
\acro{MCU}{micro-controller unit}
\acro{MHA}{multi-head attention}
\acro{MitM}{man-in-the-middle}
\acro{ML}{machine learning}
\acro{MLM}{masked language modeling}
\acro{MLP}{multilayer perceptron}
\acro{MQTT}{message queuing telemetry  transport}
\acro{NAT}{network address translation}
\acro{NIDS}{network intrusion detection system}
\acro{NLP}{natural language processing}
\acro{P2SQL}{prompt-to-SQL}
\acro{PCA}{principal component analysis}
\acro{PCAP}{packet capture}
\acro{PDF}{probability density function}
\acro{Pre}{precision}
\acro{R2L}{remote-to-local}
\acro{RCE}{remote code execution}
\acro{Rec}{recall}
\acro{ReLU}{rectified linear unit}
\acro{RL}{reinforcement learning}
\acro{RNN}{recurrent neural network}
\acro{SCADA}{supervisory control and data acquisition}
\acro{SDN}{software-defined network}
\acro{seq2seq}{sequence-to-sequence}
\acro{SHAP}{Shapley additive explanations}
\acro{SIEM}{information and event management}
\acro{SMOTE}{synthetic minority oversampling technique}
\acro{SLM}{small language model}
\acro{SMTP}{Simple Mail Transfer Protocol}
\acro{SNMP}{Simple Network Management Protocol}
\acro{SOTA}{state-of-the-art}
\acro{SQL}{Structured Query Language}
\acro{SwinT}{shifted windows Transformer}
\acro{TCP}{Transmission Control Protocol}
\acro{TL}{transfer learning}
\acro{TLS}{transport layer security}
\acro{TNN-IDS}{Transformer neural network-based IDS}
\acro{TTL}{time-to-live}
\acro{U2R}{user-to-root}
\acro{UAV}{unmanned aerial vehicles}
\acro{UDP}{User Datagram Protocol}
\acro{URL}{uniform resource locator}
\acro{URLLC}{ultra-reliable low latency communications}
\acro{V2C}{vehicles-to-cloud}
\acro{V2X}{vehicle-to-everything}
\acro{ViT}{vision Transformer}
\acro{WAF}{web application firewall}
\acro{XAI}{explainable AI}
\acro{XGBoost}{feature screening strategy based on eXtreme gradient boosting}
\acro{XSS}{cross-site scripting}
\end{acronym}

\end{multicols}
\end{table}

However, despite these advancements, current \ac{ML} or \ac{DL}-based \ac{IDS} methods still encounter new challenges, as long as network technologies evolving, such as vulnerability to attacks on central entities, decreased system performance with larger user bases, and inefficiencies in traditional centralized processing \cite{zhou2024network}. Additionally, in real-world scenarios, the attack sample data generated by an organization's or enterprise's network system tends to be of lower quality for training purposes. Consequently, the \ac{IDS} models are limited in their detection capabilities based on this data. Hence, there is a critical necessity to develop an efficient \ac{IDS} method \cite{wang2021ddostc}. Despite its strong adaptability, \ac{ML} or \ac{DL}-based exhibits high \acp{FPR} in domains where the feature distributions of malicious flows overlap with those of benign flows. Many researchers hypothesize that the limitations of conventional AI-based IDS algorithms stem from their utilization of individual flows as input data, restricting the classifier's ability to model feature distributions within a single flow \cite{nguyen2022flow}. They claim that this constraint can be overcome by employing sequences of flows, enabling the classifier to better capture the distribution of a flow relative to others. 


Current approaches detect pattern changes induced by attacks by learning the normal pattern of a sequence, typically extracted solely from network data frames like log files and host \acp{IDS}. Consequently, if the target sequence for detection contains only a minimal number of attacks, the deviation from the normal pattern will be subtle, posing challenges for detection. Therefore, there is a need for a new detection method to address such scenarios. \ac{RNN} and \ac{LSTM} models are commonly employed to grasp time series characteristics \cite{djeffal2023noise,kheddar2024deepSteg}, {it is pivotal to recognize that these models excel in capturing time dependencies within sequential data, a critical feature for log file analysis. Unlike traditional static data analysis techniques, \acp{RNN} and \acp{LSTM} can process continuous streams of data, adjusting their internal state based on the latest inputs. This dynamic adjustment is essential for detecting anomalies in log files where patterns evolve and new attack vectors emerge over time. For instance, \acp{LSTM} specifically address the challenge of long-term dependencies in sequences—an area where earlier neural models faltered. By effectively remembering information for long periods, \acp{LSTM} can detect complex attack patterns that unfold slowly, which are common in sophisticated network breaches. This capability makes them particularly valuable in scenarios where attackers employ low-and-slow tactics to evade detection. However, despite their strengths, } they suffer from decreased performance as sequences lengthen. Given the inherently sequential nature of network traffic data, new sequential \ac{AI} models, like the Transformer network model  \cite{vaswani2017attention,djeffal2023automatic}, present a logical solution to this issue.

Transformers and their variants, leveraging self-attention mechanisms \cite{vaswani2017attention}, have excelled in \ac{NLP} tasks, including text classification, dialogue recognition, and machine translation. {Turning to the application of Transformer-based models such as \ac{CNN} Transformer and \ac{ViT} in network data analysis, these models bring the advantage of deep contextual understanding. Unlike \acp{RNN} and \acp{LSTM}, which process data linearly, Transformers analyze entire blocks of text in one go, capturing intricate patterns and dependencies across much longer sequences than was previously possible. This ability stems from their attention mechanisms, which weigh the importance of each word in the context of all others in the sequence.} Inspired by the Transformer’s prowess in handling ordered data sequences, researchers have explored its applicability in intrusion and anomaly detection, demonstrating its effectiveness in various scenarios. Since network intrusion typically unfolds over time, most existing models lack the capability to capture time-series features, resulting in information loss. The attention mechanism of the Transformer, however, can effectively learn the temporal correlation of network intrusion data, thereby enhancing the accuracy of \ac{NIDS}. The potential application of these advancements lies in their utilization with various cyber-data, including computer logs and network packets, which can be structured as text and conceptualized as distinct languages. A modest yet expanding body of literature has showcased successful applications of \ac{NLP} to cyber data, such as identifying red team attacks within authentication logs. Once attacks are accurately identified, straightforward responses like quarantining or shutting down affected systems become feasible. Furthermore, leveraging windows event logs enables each computer to autonomously execute its response without necessitating feedback from other network sources. This autonomy is especially valuable in tactical networks, where connections may be unreliable and exhibit low throughput \cite{steverson2021cyber}. 

 Transformers are a main component of \acp{LLM} models due to their ability to capture complex dependencies in sequential data. For example, \ac{GPT} \cite{floridi2020gpt} and \ac{BERT} models \cite{djeffal2023automatic} use Transformers to generate coherent and contextually relevant text. By leveraging attention mechanisms, Transformers enable \ac{LLM} systems to understand and manipulate data at a more nuanced level, leading to more realistic and diverse outputs in tasks such as text generation, image generation, and speech synthesis \cite{kheddar2024automatic}. {The deployment of \ac{LLM} in \ac{IDS} leverages these contextual insights to a greater extent. By training on vast datasets, \acp{LLM} develop a nuanced understanding of what constitutes normal and anomalous network behavior. This \ac{DL} capability enables them to identify even the most subtle signs of network intrusions, often before they fully manifest, thus providing a proactive rather than reactive security measure. In practical terms, this means that \ac{GPT} and \ac{BERT} can discern subtle anomalies in network traffic and log files by evaluating the data within the broader context of network behavior. For example, \ac{GPT}’s generative capabilities allow it to predict potential future sequences based on past data, offering insights into expected versus actual network activity. Similarly, \ac{BERT}’s bidirectional approach provides a holistic view of both past and future contexts within a dataset, enhancing its ability to pinpoint anomalies that would otherwise go unnoticed in a solely forward-feeding analysis.} Thus, the \ac{LLM} could be adapted at learning the benign patterns found within network traffic sequences, making it well-suited to detect alterations in network traffic patterns induced by even a minimal number of attacks \cite{nam2021intrusion}.

This survey aims to explore the cutting-edge \ac{IDS} based on Transformers and \ac{LLM} advances in \ac{AI} techniques, which offer promising alternatives to address the weaknesses of existing \ac{ML} and \ac{DL}-based IDS \ac{SOTA} strategies. By reading this survey, the author aims to assist researchers in understanding the concept of enhancing \ac{IDS} with Transformers and \acp{LLM}, as well as shedding light on future directions and perspectives.

\subsection{Existing reviews and our contributions}

Employing Transformers, and specifically large \acp{LLM}, although a trending research topic, is still in its early stages. Consequently, there are few recent reviews, and most of them are preprints that are not yet published. These reviews summarize how \ac{NLP} algorithms could be exploited for more effective detection and mitigation of insider threats in cybersecurity. However, they tend to focus strongly on specific aspects: the use of generative \ac{AI} and \acp{LLM} to strengthen resilience and security, as seen in \cite{yigit2024critical,ferrag2024generative}; the mitigation of insider threats using \ac{NLP} and \ac{DL}, as discussed in \cite{alzaabi2024review}; the application of \acp{LLM} in cybersecurity tasks such as vulnerability detection and malware analysis, highlighted in \cite{xu2024large,zhang2024llms}; or examining the dual impact of \acp{LLM} on security and privacy, as explored in \cite{yao2024survey}. These reviews are generic and cover only a few works in each area of cybersecurity. In contrast, our survey provides a comprehensive review of empowering \ac{IDS} using Transformers, which are the main component of \acp{LLM}, along with some pre-trained \acp{LLM} models such as \ac{GPT} and \ac{BERT}.  Moreover, this survey may assist researchers in building their own \acp{LLM} tailored to intrusion detection by providing possible configurations of the Attention layer and existing Transformer and \acp{LLM} models.  The key contributions of this survey can be summarized as follows:

\begin{itemize}
    \item The survey delves into the background of \ac{IDS} and covers various types of attacks. It also provides an in-depth taxonomy of the different \ac{IDS} techniques, including \ac{HIDS} and \ac{NIDS}, employed to secure diverse environments.

    \item The survey offers an extensive taxonomy of different Transformer models applicable to strengthen various \ac{IDS} techniques, including Attention, CNN/LSTM-Transformer, \ac{ViT}, GAN-Transformer, \ac{GPT}, \ac{BERT} and their variants, among others.
    
    \item It presents a detailed classification of various Attention mechanisms and their possible configurations with traditional \ac{DL} models, such as \ac{CNN}.
    
    \item The survey reviews the latest \ac{SOTA} schemes proposed for different environments and applications, such as computer networks, the \ac{IoT} and \ac{IIoT}, critical infrastructure, cloud and \ac{SDN}, and \acp{AV}.
    
    \item Finally, the survey highlights existing research challenges and suggests potential future directions for the field of Transformers and \ac{LLM}-based IDS.
\end{itemize}

Table \ref{tab:relatedWork} provides a summary comparison of our survey with existing \ac{LLM}-based cybersecurity surveys, focusing on \textbf{\textit{content}} and survey \textbf{\textit{structure}}. It is evident that our survey thoroughly addresses all relevant areas of Transformers and \ac{LLM}-based \ac{IDS}, including background, metrics and datasets, existing Transformers and \acp{LLM} models used in \ac{IDS}, applications of \ac{LLM}-based \ac{IDS}, research gaps, challenges, and future directions. In contrast, most other surveys either do not cover these areas or only partially address them. Consequently, our survey can be considered comprehensive in its coverage of the various fields related to Transformers and \ac{LLM}-based \ac{IDS} research.

\renewcommand{\arraystretch}{1.5} 
\begin{table}[h!]
\scriptsize
\caption{Comparison with existing Transformers and LLM-based IDS. The markers \CIRCLE{}, \LEFTcircle{}, and \Circle{} denote that the particular field has been attended to, partly addressed, and overlooked, respectively.  }
\begin{minipage}{1\textwidth}
\centering
\label{tab:relatedWork}
\begin{tabular}{m{0.3cm}m{1cm}m{1cm}m{0.7cm}m{5.5cm}m{7cm}}
\toprule
CT & Reference & Year & NRTLIP  & Description of the similarity with our survey & Distinction from our survey   \\ \hline
 & \cite{yigit2024critical} & 2024 & 0 & Introducing advanced strategies utilizing Generative AI and \acp{LLM} to bolster resilience and security in critical infrastructure & Focuses solely on cyber threats to critical infrastructure, with no discussion on intrusion detection or Transformers and their related work. \\ \cline{2-6}

& \cite{ferrag2024generative} & 2024 & 10 & The review offers a comprehensive analysis of the use of Generative AI and \acp{LLM} in cybersecurity. & Focus on software and hardware general threat. A few discussions on intrusion detection
or Transformers and their related work. \\\cline{2-6}
  
\multirow{5}{*}{\rotatebox{90}{\textbf{Content}}} & \cite{alzaabi2024review} & 2024 & 2 & Recommends utilizing \ac{DL}  and \ac{NLP} for more effective detection and mitigation of insider threats in cybersecurity, highlighting the importance of incorporating time-series techniques. & It has a specific emphasis on traditional \ac{ML} methods and their shortcomings in adequately handling the complexities of insider threats. Few works related to \ac{IDS} have been reviewed.\\\cline{2-6}
 
&  \cite{xu2024large} & 2024 & 10 & Emphasizes the various applications of \acp{LLM} in cybersecurity tasks, including vulnerability detection, malware analysis, and the detection of intrusions and phishing attempts. & Discussed few \ac{IDS} techniques that utilize \acp{LLM}. It primary focus on software and system security, blockchain, and hardware. \\\cline{2-6}

& \cite{zhang2024llms} & 2024 & 17 & This paper presents a systematic literature review of many studies on the application of \acp{LLM} in cybersecurity & Covers a wide range of \ac{LLM} applications in cybersecurity, with only a few \ac{IDS} techniques reviewed. No \ac{IDS} work related to Transformers is included. \\\cline{2-6}

& \cite{yao2024survey} & 2024 & 3 & Examines the dual impact of \acp{LLM} on security and privacy, emphasizing their ability to improve cybersecurity and data protection while also introducing new risks and vulnerabilities. & Examine the positive impacts of \acp{LLM} on security and privacy, the potential risks and threats associated with their use, and the inherent vulnerabilities within \acp{LLM}. A few \ac{IDS} papers are discussed. \\\cline{2-6}

&  Ours, 2024 & 2024 & 103 & The proposed work reviews GPT, BERT, and their derivatives  \acp{LLM}, in the context of cybersecurity and their applications. & Our survey uniquely focuses on IDS schemes using Transformers (Attention mechanisms, CNN/LSTM, ViT, GAN, \acp{LLM} like GPT and BERT), providing comprehensive coverage of IDS, \acp{LLM}, metrics, datasets, and Transformer and LLM-based applications. \\
\bottomrule
\vspace{0.1cm}
\end{tabular}
\end{minipage}\hfill
\begin{minipage}{1\textwidth}
\centering
\tiny
\begin{tabular}{cccccccccccccccc}
\hline
 & Ref. & Background & Bibliometrics & \multicolumn{4}{c}{Transformers models-based IDS} & & \multicolumn{2}{c}{LLMs-based IDS} & Preproc. & Datasets & Metrics & Applications & Gaps \& FD \\ \cline{5-8} \cline{10-11}
 &  & (IDS, LLM) & analysis & Attention & CNN/LSTM & ViT & GAN & & GPT & BERT & & (IDS) & & &  \\
 \hline
 \multirow{6}{*}{\rotatebox{90}{\textbf{\scriptsize{Structure}}}} & \cite{yigit2024critical} & \LEFTcircle{} & \Circle{} & \Circle{} & \Circle{} & \Circle{} & \Circle{} &  & \Circle{} & \Circle{} & \Circle{} & \LEFTcircle{} & \LEFTcircle{} & \Circle{} & \LEFTcircle{} \\
& \cite{ferrag2024generative}  & \LEFTcircle{} & \Circle{} & \Circle{} & \LEFTcircle{} & \Circle{} & \Circle{} &  & \CIRCLE{} & \LEFTcircle{} & \Circle{} & \Circle{} & \Circle{} & \Circle{} & \LEFTcircle{} \\
 & \cite{alzaabi2024review}  & \LEFTcircle{} & \Circle{} & \Circle{} & \Circle{} & \Circle{} & \Circle{} &  & \Circle{} & \Circle{} & \Circle{} & \LEFTcircle{} & \CIRCLE{} & \Circle{} & \LEFTcircle{} \\
 
 & \cite{xu2024large}   & \Circle{} & \CIRCLE{} & \Circle{} & \Circle{} & \Circle{} & \Circle{} &  & \CIRCLE{} & \LEFTcircle{} & \CIRCLE{} & \LEFTcircle{} & \Circle{} & \LEFTcircle{}  & \CIRCLE{} \\
 
& \cite{zhang2024llms}  & \LEFTcircle{} & \CIRCLE{} & \Circle{} & \Circle{} & \Circle{} & \Circle{} &  & \CIRCLE{} & \CIRCLE{} & \Circle{} & \Circle{} & \Circle{} & \Circle{} & \CIRCLE{} \\

& \cite{yao2024survey}  & \LEFTcircle{} & \CIRCLE{} & \Circle{} & \Circle{} & \Circle{} & \Circle{} &  & \CIRCLE{} & \LEFTcircle{} & \Circle{} & \Circle{} & \Circle{} & \Circle{} & \LEFTcircle{}\\

& Our & \CIRCLE{} & \CIRCLE{} & \CIRCLE{} & \CIRCLE{} & \CIRCLE{} & \CIRCLE{} &  & \CIRCLE{} & \CIRCLE{} & \CIRCLE{} & \CIRCLE{} & \CIRCLE{} & \CIRCLE{} & \CIRCLE{} \\
 \bottomrule
\end{tabular}
\end{minipage}
\begin{flushleft}
 Abbreviations:  multi-modal (MM), future direction (FD), number of reviewed
Transformer-and LLM-based IDS paper (NRTLIP)  
\end{flushleft}
\end{table}

\subsection{Research questions and objectives}

To streamline this survey, the author  defined five research questions in Table \ref{RQs}. By following the study’s roadmap, readers will grasp the key insights and comprehend the study’s objectives. The table offers a structured summary of the research questions (RQ) and their associated motivating factors. Each row in the table addresses a specific research question, offering a concise view of the goals guiding research in automated technologies for Transformers and \acp{LLM}-based \ac{IDS}.

\begin{table*}[h!]
\caption{{Research questions for IDS utilizing Transformers and \acp{LLM}.}}
\label{RQs}
\scriptsize
\begin{tabular}{
m{5mm}
m{80mm}
m{80mm}
}
\hline

{RQ\#} & {Question}	& {Objective}   \\

\hline

{RQ1} & {What drives the utilization of Transformers and \acp{LLM} in \ac{IDS}, and what benefits do they provide compared to traditional \ac{ML}/\ac{DL} models?}
& {Gain insight into why Transformers and \acp{LLM} are employed in \ac{IDS} and distinguish their distinct advantages over conventional \ac{ML}/\ac{DL} models.} \\ \hline

{RQ2} & {What are the current methodologies employing Transformers and \acp{LLM} in \ac{IDS} for various attack types? How successful are these approaches in identifying intrusions?}	& {Identify and comprehend cutting-edge techniques employing Transformers and \acp{LLM} in \ac{IDS} across a spectrum of attack types.}  \\ \hline

{RQ3} & {How do \ac{IDS} methods based on Transformers and \acp{LLM} compare with each other and with traditional methods in terms of accuracy and efficiency in detecting intrusions?}	& {Compare and evaluate the performance of various \ac{IDS} methods based on Transformers and \acp{LLM}, contrasting them with traditional techniques.}   \\ \hline

{RQ4} & {How can we improve the interpretability of IDS models based on Transformers and \acp{LLM}? What challenges exist, and what are the current research trends in enhancing their explainability?}	& Evaluate the transparency of \ac{IDS} models based on Transformers and \acp{LLM}, outline interpretability challenges, and examine ongoing research aimed at improving their clarity.  \\ \hline

{RQ5} & {What are the primary applications of \ac{IDS} employing Transformers and \acp{LLM}  in diverse network settings?} & {Understand the practical implementations and applications of \ac{IDS} utilizing Transformers and \acp{LLM} in diverse network environments.}   \\ \hline

{RQ6} & {What are the critical areas needing additional research in Transformer and LLM-based \ac{IDS}? What potential advancements could emerge in this domain?} 
 & {Identify future potential advancements and essential areas for further research in \ac{IDS} utilizing Transformers and \acp{LLM}.}
  \\ 

\hline

\end{tabular}
\end{table*}

\subsection{Survey methodology} 

To identify and review existing studies on Transformers and LLM-based IDS, a comprehensive search was conducted across several leading publication databases renowned for their high-quality scientific research. The primary search was carried out in Scopus, which systematically includes databases such as Web of Science, Elsevier, IEEE, ACM Digital Library, Wiley, and IET Digital Library, among others.

Articles published between 2017 and 2024 were given priority. However, older publications were also considered when necessary to provide a historical context, dataset, metrics, etc. The survey focused on computer science and engineering studies from databases including IEEE Xplore, ScienceDirect, Wiley, Springer, and Taylor \& Francis. Additionally, considering the recent and trending nature of the topic, high-quality pre-prints from arXiv, SSRN, and TechRxiv were selected. Only articles written in English were included in the final analysis. Key search terms used in the \textit{abstract}, \textit{article title}, and \textit{keywords}. {To efficiently manage the search and selection of relevant papers, the following query was structured to capture articles with focused relevance to Transformers and \ac{LLM}-based IDS:}

\begin{center}
Selected papers = FROM ("Abstract" \textbar\textbar ~"Title" \textbar\textbar ~"keywords") SELECT ( References WHERE keywords = (Transformer \textbar\textbar ~LLM \textbar\textbar ~NLP ) \& ( Cyberthreat \textbar\textbar ~IDS ))
\end{center}

The symbols \& and \textbar\textbar ~signify the logical operations AND and OR, respectively.  {The construction of this query is instrumental in refining the selection process, ensuring that only the most relevant studies are considered. This targeted approach has necessitated the development of explicit inclusion and exclusion criteria, detailed as follows:}
\vspace{0.1cm}

\noindent{\textbf{Explicit inclusion criteria:}}
\begin{itemize}
    \item {Research explicitly discussing Transformers or LLMs in the context of NLP, Cyberthreat, or IDS.}
    \item {Studies must be published in indexed journals or conferences, ensuring they meet established academic standards and visibility.}

\end{itemize}

\noindent{\textbf{Explicit exclusion criteria:}}
\begin{itemize}
    \item {Studies that do not focus on the specified technological applications or topics.}
    \item {Research ambiguously referencing "Transformers" in non-computational contexts, such as electrical engineering.}
\end{itemize}

\begin{figure}[h!]
    \centering
\includegraphics[scale=0.75]{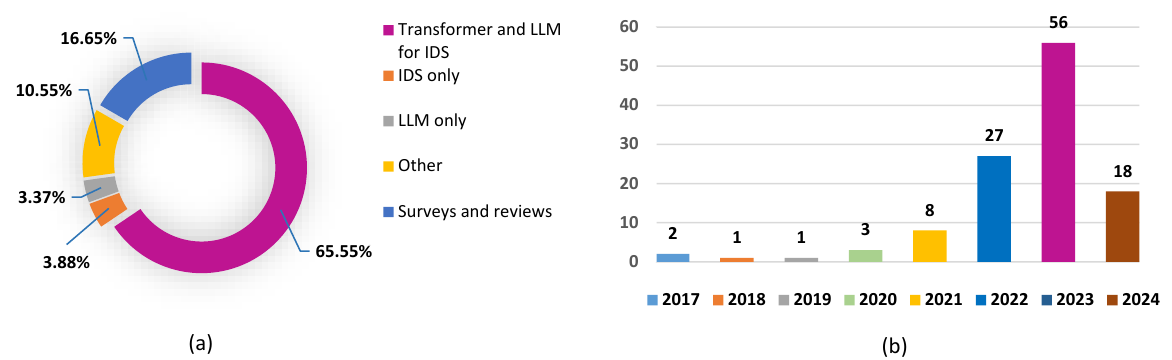}
    \caption{Bibliographic Statistics: (a) Analysis of research field and article type distribution from all references (180 papers). (b) Statistical distribution of 118 research papers on Transformers and \acp{LLM} for IDS (2017-2024).}
    \label{fig:statistics}
\end{figure}

{These inclusion and exclusion criteria are aimed at reinforcing the scientific rigor of the review process, ensuring that only pertinent and high-quality studies are considered.} The final number of selected papers on \ac{IDS} based purely on Transformers and \acp{LLM} is 118. Figure \ref{fig:statistics} provides more details about the involved research papers in terms of paper type, domain conducted, and distribution of papers based purely on Transformers and \acp{LLM}. It is obvious that Transformers and LLM-based IDS has emerged as a recent and rapidly growing field, with a notable surge in publications starting from 2022. Moreover, the inclusion of 40 pre-prints among the 180 cited papers underscores the trending nature and active research pursuit in this domain.

{To assess the quality and impact of academic work, several metrics are considered in writing the survey: H-index, Q1 ranking, impact factor of the journal, average citations, and database indexing of the paper. These metrics ensure broader visibility, accessibility, and credibility. Additionally, highly important papers are discussed extensively in the text and their findings are reported in the tables, while lower-quality papers are only discussed in the tables.}

\subsection{Survey organization}

\begin{figure}
    \centering
\includegraphics[scale=0.54]{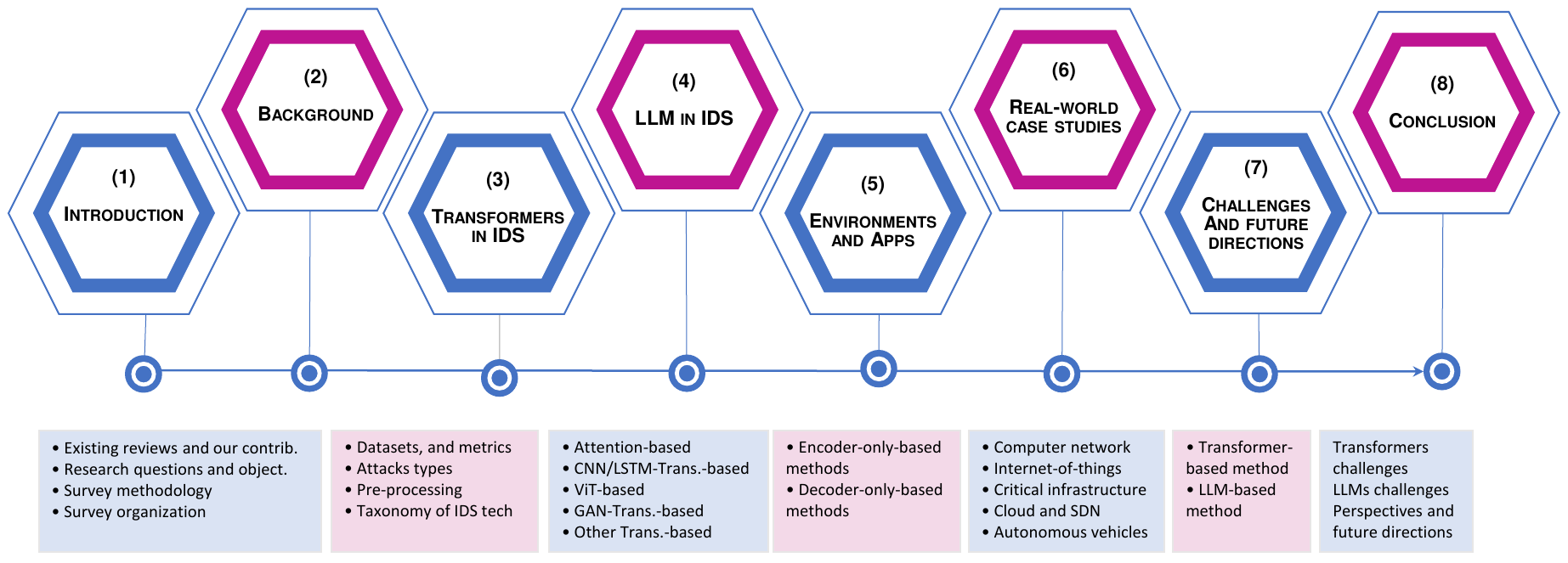}
    \caption{A roadmap outlining the primary sections and essential concepts discussed in this survey.}
    \label{fig:RM}
\end{figure}

The remainder of this survey is organized as follows: Section \ref{sec2} summarizes the essential fundamentals, including datasets, metrics, various investigated attacks, pre-processing steps, and a taxonomy of existing \ac{IDS} techniques. Section \ref{sec3} details the principles of employing Transformers in \ac{IDS} and reviews numerous related studies. Section \ref{sec4} explores the use of \acp{LLM} in \ac{IDS}, accompanied by a review of several studies. Section \ref{sec5} highlights the most significant applications investigated in the \ac{SOTA} research. Section \ref{sec6} provides real-world case studies for both Transformer-based and \ac{LLM}-based \acp{IDS} systems. Section \ref{sec7} discusses the open challenges facing Transformer and LLM-based \ac{IDS} schemes and suggests potential future contributions. Section \ref{sec8} provides a conclusion for this survey. Figure \ref{fig:RM} offers a detailed roadmap for this survey and lists the key topics covered in the study.

\section{Background}
\label{sec2}

This section covers the key elements of \ac{IDS}, starting with intrusion detection methods, followed by an overview of attack categories, the datasets used for evaluation, and the metrics employed to measure IDS performance.

\subsection{Intrusion detection}

\begin{figure}
    \centering
    \includegraphics[scale=0.8]{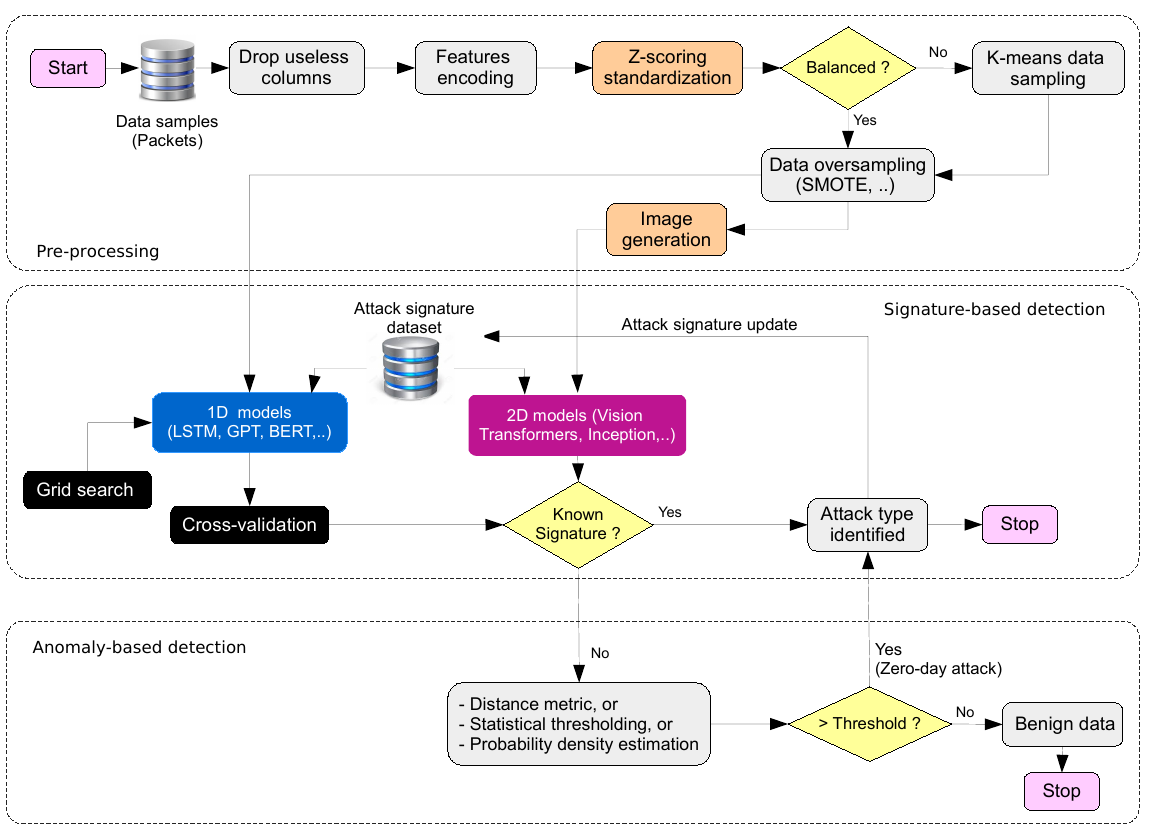}
    \caption{A standard framework diagram for a \ac{DL}-based \ac{IDS} consists of three stages: (i) preprocessing, (ii) signature-based detection, and (iii) anomaly-based detection.}
    \label{fig:IDSCat}
\end{figure}

This part elucidates the application of \ac{DL} in \ac{IDS} for detecting cyber-attacks, detailing procedures that can be applied universally across various environments. Figure \ref{fig:IDSCat} presents the overall architecture of a generic detection system, which consists of two main phases: data pre-processing and \ac{IDS} framework development. The \ac{IDS} framework can be either signature-based or anomaly-based, followed by training and testing. The following sections provide detailed information about each category.

\subsubsection{Data pre-processing}

Computational efficiency can be improved and memory usage decreased by downsizing the dataset and removing less significant features. This can be achieved by dropping some dataset's columns that are expected to be unaffected by the attacks being studied, such as \ac{MAC} addresses, timestamp, among others. Encoding feature attributes in processed data can be accomplished using various techniques, depending on the data's nature. Common methods include \textit{label encoding}, suitable for categorical data when categories are ordinal; \textit{one-hot encoding}, ideal for categorical data lacking ordinal relationships; and \textit{binary encoding}, which combines benefits from both label and one-hot encoding. These techniques offer versatile ways to represent categorical features effectively in \ac{ML} models. z-score normalization, is a technique used to standardize numerical features in a dataset. It involves subtracting the mean ($\mu$) of the feature from each data point and then dividing by the standard deviation ($\sigma$) of the feature. Mathematically, the formula for z-score normalization is represented as:

\begin{equation}
\label{Znorm}
 z = \frac{{x - \mu}}{{\sigma}}   
\end{equation}

\noindent Where, \( x \) is the original value of the feature, \( \mu \) is the mean of the feature, \( \sigma \) is the standard deviation of the feature, and  \( z \) is the standardized value, also known as the z-score. This transformation ensures that the standardized feature has a \( \mu \) of 0 and a \( \sigma \) of 1, which is particularly useful for algorithms that assume normally distributed data or require features to be on a similar scale for proper convergence. It also facilitates direct comparison and analysis, while also reducing data bias and enhancing stability and reliability. The classes of data are evaluated to determine if they are balanced or not. If necessary, k-means  undersampling \cite{zhou2023adaptive} is employed to expedite computation when data are balanced. k-means is a commonly used clustering algorithm that divides a dataset into k distinct clusters, reducing data size without losing meaningful information by minimizing the distance between data points and their cluster centers. However, k-means is sensitive to initial cluster centers and noisy data. The mini k-means algorithm \cite{ikotun2023k}, an improved version, mitigates these issues by selecting random points as initial cluster centers, assigning data points to the nearest centers, and recalculating centroids until convergence is achieved or maximum iterations are reached.  The  imbalance data could skew the training towards the more abundant samples, under-training the less represented attack types and reducing the model's overall applicability. To address this, researchers used oversampling techniques to balance the dataset, ensuring the model is adequately trained on all attack types. \Ac{SMOTE} is a widely used method to address data imbalance by generating synthetic samples for minority classes \cite{fernandez2018smote}, thus balancing the dataset. It works by selecting a sample from the minority class, finding its $k$ nearest neighbors, and generating new synthetic samples by interpolating between them. While the mentioned data pre-processing steps are common in DL-based \ac{IDS}, they are not necessarily used all at once for the same method. Additionally, the data can be converted to a 2D format to be processed as an image, allowing the application of existing 2D \ac{DL} algorithms such as \ac{ViT} and Inception pre-trained models.

\subsubsection{Taxonomy of IDS techniques}

After pre-processing, the researchers apply various \ac{DL} algorithms to build an efficient DL-based \ac{IDS} model. This phase involves model training and optimization, including cross-validation, enhanced convergence tools, and grid search techniques to assess the model's performance. The DL-based \ac{IDS} frameworks can be categorized into two types, depending on the availability of attack type information:\\

\noindent \textbf{(a) Signature-based model:} Are detection techniques in \ac{IDS}, which involve matching network traffic or system actions against a database of attack signatures to trigger alerts upon detection. It outlines the process of comparing observed data \( D \) with a signature database \( S = \{s_1, s_2, ..., s_n\} \) using a matching function \( Match(S, D) \). While effective against known attacks, this method has limitations in detecting new or mutated attacks and may not be suitable for resource-constrained environments like \ac{IoT}. Research explores pattern-based detection methods as alternatives to address these challenges. Experts evaluate the trained model's performance on test instances. If the results are unsatisfactory, techniques such as \ac{PCA} may be applied as a corrective measure to reduce the complexity of the data and enhance interpretability, thereby improving the performance of \ac{DL} algorithms. Once the results meet the accuracy criteria, the model becomes capable of effectively identifying and distinguishing various attack types. \ac{ViT} and pre-trained models such as Inception are widely used for their ability to capture intricate patterns and features from network traffic data. However, several DL algorithms are tailored for packet classification rather than 2D image representation. These include various 1D DL techniques such as \ac{LSTM}, \ac{RNN}, 1D \ac{CNN}, \ac{BERT}, and others. Tools like \textit{AutoKeras} \cite{jin2019auto} and \textit{H2O} \cite{nguyen2019machine} automate model-building, ensuring optimal architecture selection and hyperparameter tuning. This simplifies the complex task of designing an effective \ac{IDS}. To enhance the performance of DL algorithms, experts employ optimization techniques such as, but not limited to, \textit{AdamW}  and \textit{BO-TPE} \cite{tang2023intrusion}. \textit{AdamW} improves convergence by optimizing training trajectories, while \textit{BO-TPE} selects optimal hyperparameters, fine-tuning algorithm configurations.\\

\noindent \textbf{(b) Anomaly-based models:} Are detection techniques employed when signature-based IDS techniques fail to detect zero-day threats. This technique monitors normal behavior to detect deviations, alerting anomalies beyond predefined thresholds without classifying specific attacks. Techniques such as statistical thresholding utilize the mean (\(\mu\)) and standard deviation (\(\sigma\)), employing a threshold (\(\mathrm{Threshold} = \mu + k \cdot \sigma\)) to detect attacks. Distance metrics measure dissimilarities between observed ($O$) and baseline ($B$) behaviors, often using Euclidean distance. Probability density estimation, like \ac{GMM}, models normal behavior and computes anomaly scores inversely to \ac{PDF}. While ML/DL models learn normal and attack behaviors, establishing standard profiles proves more effective than solely identifying known attacks. Anomaly-based techniques excel in detecting new threats but struggle with establishing accurate baseline profiles. Once an attack is detected using anomaly-based techniques, it becomes a known attack, as these techniques update the signature dataset with the new signature of the detected threat. Anomaly-based techniques often result in higher \ac{FAR}. Research addresses this challenge by employing ML/DL to construct robust normal behavior profiles, utilizing advanced methods such as meta-classifiers and hybrid feature selection \cite{tama2019tse}.

\subsection{Attack categories, datasets, and metrics}

Various datasets in \ac{IDS} research cater to diverse network environments (e.g., network, host, \ac{ICS}, \ac{IoT}, cloud, edge, smart grid), encompassing a wide array of attack types categorized as follows:

\begin{itemize}
    \item \textbf{\ac{DoS}:} Attacks aim to render systems or networks inaccessible. Examples include \ac{TCP}/\ac{UDP} flooding, \ac{ICMP} Smurf/ping of death, IP teardrop, UDPstorm, \ac{DNS} amplification, \ac{TCP} SYN/ACK flood, and \ac{HTTP} Slowloris.
    
    \item \textbf{\Ac{R2L}:} Remote attacks exploit vulnerabilities to gain local access, targeting protocols like \ac{FTP}, \ac{IMAP}, \ac{HTTP}, \ac{SMTP}, \ac{DNS}, X11, and \ac{SNMP}, with attacks like FTP\_write, phf, and SNMPgetattack.
    
    \item \textbf{\Ac{U2R}:} Exploits system vulnerabilities to escalate privileges, targeting protocols such as \ac{UDP}, \ac{TCP}, Perl, \ac{SQL}, and \ac{HTTP}, with attacks like buffer\_overflow, rootkit, HTTPtunnel, and SQLattack.
    \item \textbf{PROBE:} Scans networks for vulnerabilities using tools like IPsweep, nmap, Portsweep, Satan, mscan, and saint to gather information for potential future exploitation.
    
    \item \textbf{Enduring:} Persistent weaknesses in software, hardware, networks, or practices vulnerable to exploits over time. These vulnerabilities persist despite technological advancements and security measures, exploited for attacks like ransomware and phishing, demanding ongoing vigilance, updates, and robust security practices in \ac{IDS}.
    
\end{itemize}

For example, the work in \cite{nguyen2023transformer} studies the vulnerability of the \ac{CAN} protocol to various attacks, including flood attacks, fuzzy attacks, spoofing attacks, and replay attacks. Flood attacks, which fall under either \ac{DoS} or \ac{DDoS} categories. Fuzzy attacks are categorized as PROBE attacks. Spoofing attacks, which can include \ac{IP} spoofing, \ac{DNS} spoofing, email spoofing, etc., are classified based on context as either \ac{R2L} or \ac{U2R} attacks. Replay attacks, being a type of man-in-the-middle attack, fall within the \ac{R2L} category.

Generally, the intrusion attacks are either collected from real networks or generated from testbeds and stored as datasets. Our previous work in \cite{kheddar2023deep} provided comprehensive insights into numerous \ac{IDS} datasets. However, this survey introduces additional datasets specifically utilized in research involving Transformers and \ac{LLM}-based \ac{IDS}, extending the work already reviewed in this survey with further details cited herein. Table \ref{tab:datasets} offers a detailed overview of each dataset, including the number of features, duration of the attack, number of samples, attack/benign ratio, whether the dataset is labeled, suitable scenarios, related works, attack types, availability links, and the \ac{SOTA} \ac{AI} usage (Transformers or \acp{LLM}).

Various evaluation metrics have been employed to assess the performance of Transformers and \ac{LLM}-based \ac{IDS} in detecting various categories and types of attacks detailed in the datasets discussed (Table \ref{tab:datasets}).  Common metrics like \ac{Acc}, \ac{Rec}, \ac{Pre}, and \ac{F1} are frequently applied in general \ac{DL} tasks, with their definitions available in \cite{kheddar2024deepSteg,kheddar2023deepSpeech,habchi2023ai}. Below is a summary of additional metrics specifically used for evaluating Transformers and \ac{LLM}-based \ac{IDS}.

\begin{itemize}
    \item \textbf{\ac{MCC}:} is a measure of the quality of binary classifications. It considers true positives, true negatives, false positives, and false negatives, providing a balanced assessment even when classes are of very different sizes. Mathematically, the \ac{MCC} is defined as \cite{ghourabi2022security}: 
    \begin{equation}
     \mathrm{MCC = \frac{TP \cdot TN - FP  \cdot  FN}{\sqrt{(TP + FP)(TP + FN)(TN + FP)(TN + FN)}}}    
     \end{equation}

\noindent Where TP, TN, FP, and FN indicate true positives, true negatives, false positives, and false negatives, respectively. The \ac{MCC} ranges from -1 to 1, where: -1 indicates perfect disagreement between the predicted and actual classifications, 0 indicates no better than random classification, and 1 indicates perfect agreement between the predicted and actual classifications.

\item \textbf{\Acf{FR}:} is an essential measure for evaluating the effectiveness of adversarial attacks. It quantifies the proportion of data samples that experience a shift in the model's predicted label following adversarial manipulation (as shown in Equation 7). This metric holds significant importance in assessing adversarial attacks, especially in targeted scenarios, where it indicates the percentage of samples successfully misclassified as the desired target label.

\begin{equation}
\mathrm{FR = \frac{\text{Number of samples with changed predictions}}{\text{Total number of adversarial samples}}}
\end{equation}

\item \textbf{\Acf{AS}:} the baseline scenario represents the "normal activity" utilized to compute the alert scores. This scenario does not involve any attacks and was not included in the training data. The alert score for each log is determined as follows:

\begin{equation}
\text{AS} = \frac{\text{Score} - \text{Max baseline score}}{\text{Baseline standard deviation}}   
\end{equation}

The \textit{Score} is the anomaly score of the log being evaluated, while the \textit{Max baseline score} and \textit{Baseline standard deviation} come from the baseline scenario's anomaly scores. An alert score of one indicates the log is one standard deviation above the maximum baseline score. The threshold for detecting attacks can be adjusted according to operational needs: \textit{higher} to reduce false positives, and \textit{lower} to reduce false negatives. This threshold can also be fine-tuned over time based on practical experience and the specific network environment \cite{steverson2021cyber}.

\end{itemize}

\begin{table}[!t]
\scriptsize
\caption{Analyzing current Internet datasets for the evaluation of Transformers and \acp{LLM}-based IDSs capabilities. The symbol \CIRCLE{} indicates that a specific AI algorithm has utilized the dataset, while \Circle{} indicates that no AI algorithm has used the dataset.}
\label{tab:datasets}
\centering
\begin{tabular}
{m{2cm}m{0.2cm}m{1cm}m{1cm}m{2.5cm}m{0.2cm}m{1.5cm}m{0.5mm}m{0.5cm}m{0.5cm}m{0.5cm}m{0.5cm}m{0.2mm}m{1cm}m{0.5cm}}
\toprule
 \multicolumn{8}{c}{ Characteristics} & \multicolumn{5}{c}{ Initial usage of the dataset} &  \multicolumn{2}{c}{ Availability} \\
 \cline{2-7} 
 \cline{9-12}
 \cline{14-15}
 Dataset  &  F &  S/DoA &  ABR ($\approx$) &   Attacks types &  L? &  SS & & BERT & GPT & Transf. & LSTM & & UI &  Link \\
\hline
5GC PFCP \cite{amponis20235g}    & 36 & 16h    &--           & DoS  & Yes & 5G core IDS & & \Circle{}  & \Circle{} & \Circle{}  & \CIRCLE{}  & & \cite{tian2023adseq,pell2024lstm} & Yes\tablefootnote{\url{https://zenodo.org/records/7888347\#.ZFejbNJBxhE}} \\\hline

In-Vehicle   & 4 & 8.69 M    & 1:11           & Flooding, fuzzy, Spoofing, Replay & Yes & AV IDS & & \CIRCLE{}   & \Circle{} & \Circle{}  & \Circle{}  & & \cite{alkhatib2022can,fu2024iov}  & Yes\tablefootnote{\url{https://ieee-dataport.org/open-access/car-hacking-attack-defense-challenge-2020-dataset}} \\\hline

MQTTset  \cite{vaccari2020mqttset}     & 33 & 331 K    & 1:1           & Legitimate, 
slowite, bruteforce, malformed data, flooding, DoS  & Yes & IoT IDS& & \Circle{}  & \Circle{} & \CIRCLE{}  & \CIRCLE{}   & &\cite{wang2023res} & Yes\tablefootnote{\url{https://www.kaggle.com/datasets/cnrieiit/mqttset}} \\\hline

OTIDS   & 4 & 4.6 M    & 1:1           & DoS, Fuzzy,  Impersonation & Yes & AV IDS & & \CIRCLE{}   & \Circle{} & \Circle{}  & \Circle{} & & \cite{nwafor2022canbert} & Yes\tablefootnote{\url{https://ocslab.hksecurity.net/Dataset/CAN-intrusion-dataset}}\\\hline

NAB \cite{ahmad2017unsupervised}   & -- & 0.36 M   & 1:315           & Spatial and temporal anomalies & Yes & Time Series Anomaly IDS & & \Circle{}  & \Circle{} & \CIRCLE{}  & \Circle{}  & &\cite{li2021dct} & Yes\tablefootnote{\url{https://github.com/numenta/NAB}} \\\hline

BGL \cite{he2016experience}   & 2 & 4.7 M    & 1:13           & Failures, anomalies & No & Log anomaly IDS & & \CIRCLE{}  & \CIRCLE{} & \Circle{}  & \Circle{} & & \cite{balasubramanian2023transformer} & Yes\tablefootnote{\url{https://github.com/logpai/loghub}} \\\hline

Car-Hacking \cite{song2020vehicle} & 1 & 16.5 M    & 1:6           & DoS, fuzzy, spoofing gear, spoofing RPM & Yes & AV IDS & & \Circle{}  & \Circle{} & \CIRCLE{} & \Circle{}  & &\cite{nguyen2023transformer,abdel2021federated} & Yes\tablefootnote{\url{https://ocslab.hksecurity.net/Datasets/car-hacking-dataset}} \\\hline
NVD   & 8 & 28 K    & --        & Software vulnerabilities & Yes &  Prioritization and attacks prediction  & &  \CIRCLE{}  &  \CIRCLE{}  & \Circle{}  & \Circle{} & & \cite{aghaei2023automated} & Yes\tablefootnote{\url{https://nvd.nist.gov/}} \\\hline
ECU-IoHT & 4 & 11.2 K    & 4:1           & Smurf, Nmap Port Scan, ARP Spoofing, and DoS & Yes & Cyberattacks on Internet of health things & & \CIRCLE{}  &  \Circle{} & \Circle{}  & \CIRCLE{}  & & \cite{ghourabi2022security} & Yes\tablefootnote{\url{https://www.sciencedirect.com/science/article/abs/pii/S1570870521001475}} \\\hline

CIDDS-001large  & 92 & 28 M    & 1:9           & DoS, PortScan, PingScan, BruteForce & Yes & Flow network intrusion  & & \CIRCLE{}  &  \Circle{} & \Circle{}  & \Circle{} & & \cite{nguyen2022flow} & Yes\tablefootnote{\url{https://github.com/markusring/CIDDS?tab=readme-ov-file}} \\\hline

CSIC 2010 & 31 & 97 K    & 2:3           & Zed attack proxy, w3af, request typo errors  & No & Web attacks & & \Circle{}  & \Circle{} & \CIRCLE{} & \Circle{}   & & \cite{deshpande2023weighted} & Yes\tablefootnote{\url{https://www.kaggle.com/datasets/ispangler/csic-2010-web-application-attacks}} \\\hline

HDFS  & -- & 11 M    & 1:663           & Failures, anomalies & No & Log anomaly
IDS & & \Circle{}  & \Circle{} & \CIRCLE{} & \Circle{} & & \cite{unal2022anomalyadapters} & Yes\tablefootnote{\url{https://github.com/logpai/loghub/tree/master/HDFS}} \\\hline
SMD \cite{su2019robust}  & -- & 1.4 M    & 1:25          & Failures, anomalies & Yes &  Internet server   & & \Circle{}  & \Circle{} & \CIRCLE{} & \Circle{} & & \cite{li2022anomaly} & Yes\tablefootnote{\url{https://github.com/NetManAIOps/OmniAnomaly}} \\\hline
\\\hline
X-IIoTID \cite{al2021x} & 68 & 821 K    & 1:1           & Scanning,  websocket fuzzing, discovering, Brute force, ..etc. & Yes & IIoT IDS & & \Circle{}  & \Circle{} & \CIRCLE{} & \Circle{}  &  &\cite{chai2023ctsf}& Yes\tablefootnote{\url{https://ieee-dataport.org/documents/x-iiotid-connectivity-and-device-agnostic-intrusion-dataset-industrial-internet-things}} \\\hline

WUSTL-IIoT-2021  & 41 & 1.19 M    & 1:13           & Command injection, DoS, reconnaissance, backdoor & Yes & IIoT IDS & & \Circle{}  & \Circle{} & \CIRCLE{} & \Circle{} & & \cite{casajus2023anomaly} & Yes\tablefootnote{\url{https://www.cse.wustl.edu/~jain/iiot2/index}} \\\hline

syscall\_args  & 64 & 576 K    & --           & Web requests & Yes & Web IDS & &  \Circle{}  & \Circle{} & \CIRCLE{}  & \CIRCLE{}  & & \cite{fournier2021improving} &  Yes\tablefootnote{\url{https://zenodo.org/record/4091287\#.X4hhGNjpNQI}} \\\hline

{Web server log} & {9} & {37,693} & {1:6} & {Suspicious and dangerous activities}  & {Yes} & {Web IDS} & & {\CIRCLE{}} & {\CIRCLE{}} & {\Circle{}}   & {\Circle{}}   & & {\cite{karlsen2024large} } & {Yes\tablefootnote{\url{https://ieee-dataport.org/open-access/apache-web-server-access-log-pre-processing-web-intrusion-detection}}} \\
\bottomrule
\end{tabular}
\begin{flushleft}
\scriptsize{
  Abbreviations: Features (F);  duration of the attack (DoA); Samples (S); attack/benign ratio (ABR); labeled (L); suitable scenarios (SS); multi-stage attacks (MSA); anonymized Internet traffic (AIT); Used in (UI).}
\end{flushleft}
\end{table}

\section{{Transformers-based IDS methods}}
\label{sec3}

This section categorizes the usage of Transformers in \ac{IDS} based on their detection approaches, including Attention-based methods, CNN/LSTM-Transformer methods, \ac{ViT} methods, GAN-Transformer methods, and \ac{LLM} methods such as GPT-based and BERT-based approaches. Table \ref{tab:llm_compare} provides a comprehensive comparison between Transformers and \acp{LLM}, detailing their suitability for various problem scenarios, as well as highlighting the advantages and disadvantages associated with each algorithm. Table \ref{table:ResFind} summarizes various research findings on Transformer-based IDS schemes, including the dedicated tasks, comparison methods, datasets used, obtained results, and improvements achieved.

\begin{table*}[htbp]
\centering
\scriptsize
\caption{Comparison of various Transformers and LLM models applied for IDS field.}
\begin{tabular}{m{2cm}m{4cm}m{5cm}m{5cm}}
\hline
\textbf{Algorithm} & \textbf{Problem Scenario} & \textbf{Advantage} & \textbf{Disadvantage} \\
\hline
Self-Attention & Focuses on relevant parts of the input data for anomaly detection & Improves detection by focusing on the most important parts of the data, {increasing the efficacy of anomaly identification} & Can be computationally expensive and may require extensive tuning, {posing challenges in dynamic network environments} \\
\hline
Multi-head \newline Attention & Enhances self-Attention by using multiple Attention heads for better anomaly detection & Provides multiple perspectives on the data, improving detection robustness, {thus reducing false positives and negatives in IDS} & Increases model complexity and computational requirements, {requiring significant resources for optimal performance} \\
\hline
{CANINE Attention} & {Applied for deep analysis of of unsegmented sequential data packets in network traffic} & {Effectively manages diverse, unsegmented data types, significantly enhancing robustness in anomaly detection.} & {Requires extensive pre-processing of data, leading to increased operational overhead.} \\
\hline
{Local (hard) \newline Attention} & {Focuses on critical segments of data streams for quick anomaly detection in high-speed networks} & {Enhances computational efficiency by focusing on key data points} & {May miss anomalies outside the focused regions, limiting detection scope} \\
\hline
{Cross-Attention} & {Employs in multi-modal IDS systems to analyze varied data sources like logs and network flows} & {Enables thorough cross-referencing of data sources, improving detection accuracy} & {Integration complexity and multi-source data management could be cumbersome} \\
\hline
{Conditional \newline Attention} & {Utilized in scenarios where attention is modulated based on external conditions in IDS} & {Adapts dynamically to changing network environments, improving detection accuracy} & {Complex integration with existing IDS frameworks may present challenges} \\
\hline
CNN-Transformer & Used for combining spatial feature extraction with sequential modeling & Captures both local and global patterns in data, {specifically enhancing anomaly detection in IDS} & Computationally intensive due to combining two complex models, potentially slowing real-time analysis \\
\hline
Vision Transformer & Adapts vision-based Transformer models for network traffic visualization and anomaly detection & Excellent at capturing intricate patterns in high-dimensional data, {allowing for detailed anomaly visualization} & High computational cost and requires large datasets for training, {which may be impractical for many organizations} \\
\hline
GAN-Transformer & Combines \acp{GAN} with Transformers for detecting and simulating anomalies & Generates realistic anomalies, enhancing detection robustness, {making the IDS adaptive to evolving threats} & High complexity and requires extensive training time, {which may delay deployment and updates} \\
\hline
BERT & Applied for analyzing {bidirectional} network traffic data for anomaly detection & Leverages {forward and backward} processing context for a better understanding of data patterns, {enhancing precision in identifying network anomalies} & Requires large amounts of data for fine-tuning and significant computational resources, {limiting its applicability in smaller-scale environments} \\
\hline
RoBERTa & Fine-tuned for anomaly detection in network traffic using robustly optimized BERT approach & Improved performance over BERT due to enhanced training techniques, {resulting in more accurate anomaly detection} & Even higher computational requirements and data needs compared to BERT, {making it less suitable for resource-constrained settings} \\
\hline
GPT & Generates responses based on network behavior patterns for identifying anomalies & Effective in generating realistic sequences, useful for simulation and detection, {thereby aiding in proactive IDS measures} & May struggle with understanding context in bidirectional sense compared to BERT, {leading to potential gaps in detection capabilities} \\
\hline
\end{tabular}
\label{tab:llm_compare}
\end{table*}

\subsection{{Taxonomy of Attention Transformer} }

An Attention Transformer is a \ac{DL} model that processes data through a mechanism known as "Attention," allowing the model to weigh the importance of different parts of input data differently. It excels in understanding context and relationships within data, making it highly effective for \ac{NLP} tasks.

In computer security, Attention Transformers can enhance threat detection, anomaly detection, and phishing email identification by learning to recognize subtle patterns and anomalies in data that traditional methods might miss. In a simplified form, the Attention mechanism can be represented by the equation \cite{kheddar2024automatic}: 
\begin{equation}
    Attention(Q, K, V) = softmax \Big(\frac{QK^T}{\sqrt{d_k}}\Big).V
\end{equation}

\begin{figure}[h!]
    \centering
    \includegraphics[scale=0.67]{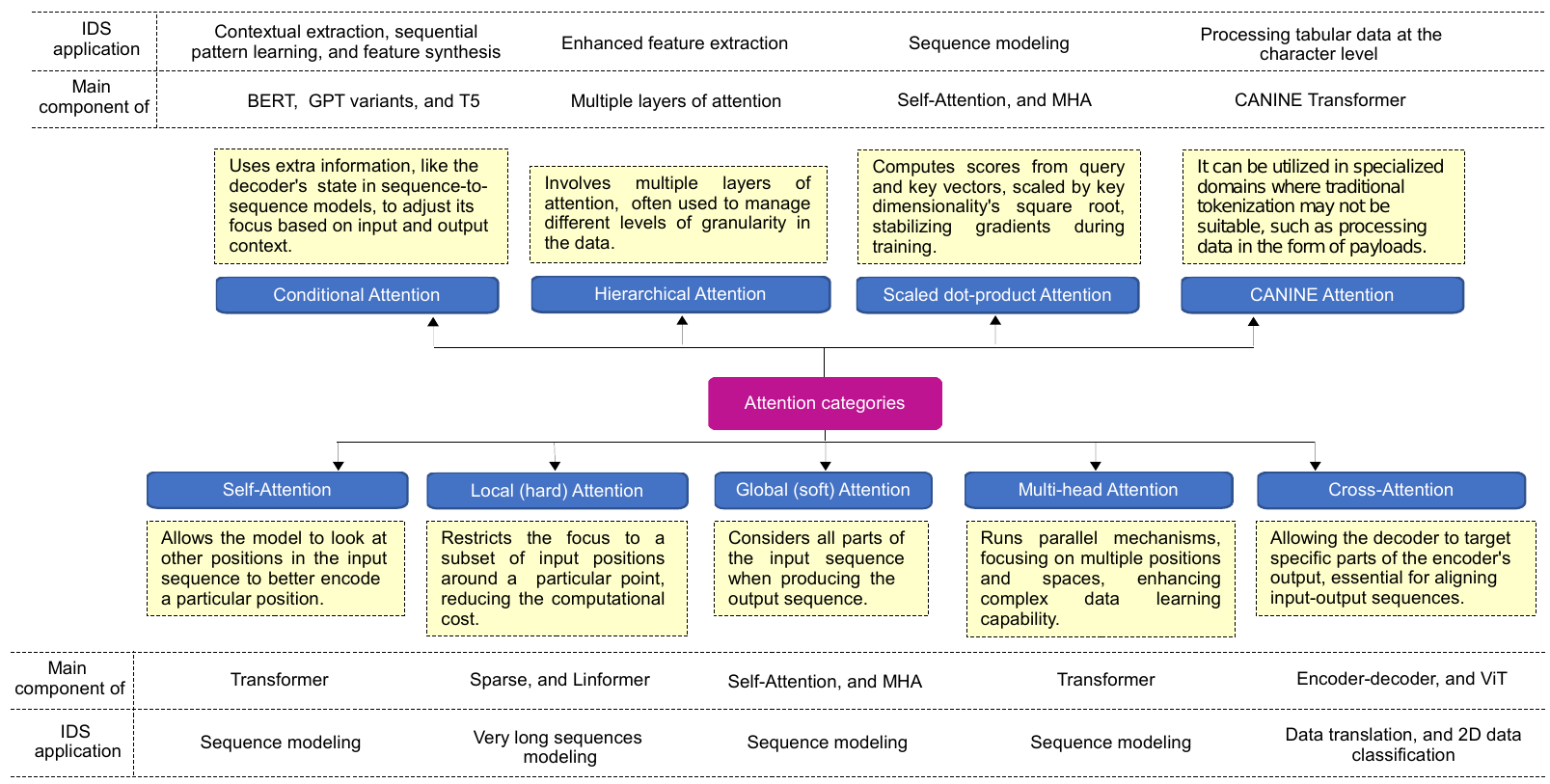}
    \caption{Visual guide to attention types in seq2seq models, illustrating how focus shifts with input and output, along with their respective related works. The primary components of self-attention (used in \cite{nguyen2023transformer,saikam2024eesnn,wang2024transformer,lan2022cascaded,wang2023robust,wang2023securing}) and its respective \ac{MHA} (used in \cite{zhou2024network, wang2021ddostc, wang2023res, abdel2021federated, hsiao2023detection, barut2022r1dit, dang2023using, lan2022cascaded, ahmed2023modified, wang2022microcontroller, li2022tcmal, liu2020deepsqli, sun2022hierarchical, han2023network, liu2023intrusion, zhang2022efficient, ullah2023tnn, wu2022rtids, sun2023intrusion, mao2022network, diaba2023scada, salam2023deep, long2024transformer, alzahrani2022anomaly, cobilean2023anomaly}) are scaled dot-product and global (soft) attentions. Hierarchical attention constructs multiple layers of attention. Local hard attention is the main component of Sparse Transformer and Linformer. Cross-attention is a key component of encoder-decoder models (used in \cite{hou2023densely}), and \ac{ViT} Transformers. Conditional attention (used in \cite{liang2023generative,moskal2023llms}) is central to many \ac{LLM} models such as \ac{BERT}, \ac{GPT} variants, and T5. CANINE attention is unique to the CANINE Transformer (used in \cite{duzgun2024network}), where tokenization principles are not adhered to or realized.  }
    \label{fig:attention}
\end{figure}

\noindent In this mechanism, \(Q\), \(K\), and \(V\) represent queries, keys, and values, respectively. Queries are used to direct the Attention mechanism's focus. Keys are part of the input data used to calculate Attention weights, indicating the relevance of different parts of the input data. Values are the actual content that, after the application of Attention weights, are aggregated to form the output of the Attention mechanism. Additionally, \(d_k\) denotes the dimension of the keys. This approach enables the model to concentrate on the most relevant parts of the data, thereby enhancing the precision and efficiency of security systems through more accurate threat identification and reduced false positives. Figure \ref{fig:attention} summarizes the existing Attention categories, {their roles in building various types of Transformers, and their related \ac{IDS} applications}.

In the architecture of many Transformer-based models, such as those used in \ac{IDS} tasks, the summation and normalization layer as described in Equation \ref{Znorm}, and the \ac{FFNN} as detailed in Equation \ref{FFNN}, often accompany the Attention layer. The \ac{FFNN} serves as a vital component for feature transformation within the Attention mechanism. It typically comprises two fully connected layers, where the first layer applies the \ac{ReLU} activation function, promoting non-linearity, and the second layer operates without an activation function. Mathematically, the \ac{FFNN} can be represented as:
\begin{equation}
\label{FFNN}
   \mathrm{FFNN(x) = ReLU(x.W_1 + b_1)W_2 + b_2 }
\end{equation}

\noindent Where, \( x.W_1 + b_1 \) is the output of the first fully connected layer in a \ac{FFNN} with weights \(W_1\) and bias \(b_1\), followed by another linear transformation with \(W_2\) and bias \(b_2\).

\subsection{Attention-based methods}

Several techniques of Attention-based \ac{IDS} have been recently proposed. For example,  \cite{hsiao2023detection} combine \ac{CNN} and \ac{GRU} along with an Attention mechanism, drawing inspiration from contemporary language models, to develop a novel and effective \ac{IDS} system to tackle SQL injection and \ac{XSS} attacks. This system is capable of reaching greater accuracy levels, requires a smaller dataset for training, and reduces the duration of the training process.

\begin{figure}[h!]
    \centering
    \includegraphics[scale=0.9]{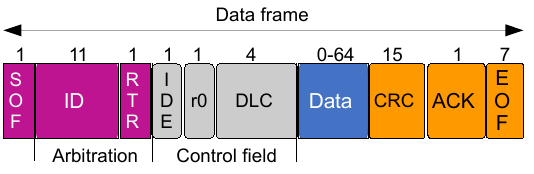}
    \caption{Format of \ac{CAN} data frame with bit-lengths for each field.}
    \label{fig:can}
\end{figure}

Self-Attention is a mechanism allowing a model to weigh the importance of different positions within the same input sequence for generating a representation. The study \cite{barut2022r1dit} introduces the residual 1-D image Transformer (R1DIT) model to address privacy and generalization issues in malware traffic classification. It parses network headers without compromising sensitive data, using \ac{DL} and \ac{MHA} mechanisms to differentiate between malware and benign traffic, enhancing privacy and adaptability to new threats like \ac{DDoS} on \ac{TLS}. \ac{RNN}-based cyber-defense approaches are limited by their sequential data processing, relying solely on the hidden state from past data, which can lead to overlooked contextual features. Nguyen et al. in their paper \cite{nguyen2023transformer}  presents a novel multi-class IDS for vehicle \ac{CAN} bus security using a Transformer-based Attention network. The fields of the \ac{CAN} bus are described in Figure \ref{fig:can}. It surpasses \ac{RNN} limitations by employing self-Attention for attack classification and replay attack detection,  through the aggregation of sequential \ac{CAN} IDs, without requiring message labeling. The model also utilizes \ac{TL} to enhance performance on small datasets from diverse car models. The work in \cite{dang2023using}, utilizes four advanced algorithms for intrusion detection: LightGBM, XGBoost, CatBoost, and a \ac{MHA} Transformer. The effectiveness of their proposed method was assessed using a well-known dataset known as CICIDS-2017. The Transformer architecture slightly surpasses LightGBM, XGBoost, and CatBoost in accuracy and efficiency, making it the preferred choice for performance. In \cite{saikam2024eesnn}, Transformer utilizing self-Attention has been used to to capture temporal characteristics in order to improve network security when facing data imbalance issue.

DNN models often produce a high number of incorrect predictions in unbalanced intrusion datasets, particularly affecting minority classes. The authors in \cite{lan2022cascaded} aim to overcome the mentioned DNN limitations through the strategic combination of \ac{DT} algorithms and feature tokenizer-Transformers based on \acp{MHA}. Initially, a \ac{DT} algorithm distinguishes between normal and malicious traffic. Then, the Transformer categorizes the malicious traffic to pinpoint the specific type of attack. Moving forward, \ac{IoT} devices face threats like data theft and DDoS attacks, leading to costly security breaches. There's a high demand for robust \ac{IDS}. Traditional models often fail to detect varied attack types due to limited adaptability. Ahmed et al. \cite{ahmed2023modified} introduce an \ac{IDS} based on a \ac{MHA}-based Transformer mechanism, showing significant improvements in accuracy, precision, recall, and F-score compared to LSTM and RNN models.

Cross-Attention enables a model to attend to different positions in another sequence, facilitating interaction between two distinct sequences for tasks like translation or summarization. In the context of cyber-security, the Denseformer model, proposed in \cite{hou2023densely}, integrates multiple Transformer-like elements into a multi-layered architecture, featuring encoder and decoder sub-layers with both self and cross-Attention mechanisms. It distinctively utilizes cross-fusion in multi-branch structures, functioning as an Attention network within dense layers to more effectively identify feature correlations. This leads to improved generalization and cyberattack detection, achieving an accuracy of 85.65\% on the NSL-KDD dataset. In \cite{wang2024transformer}, Attention has been used after the consolidation of alerts, based on a threshold, for the prediction of attacks in a multi-stage offensive.

\subsection{CNN/LSTM-Transformer-based methods}

{A CNN-Transformer combines the strengths of \acp{CNN} and Transformers, two powerful architectures in \ac{DL}. \acp{CNN} excel in processing structured grid data like images, efficiently capturing local dependencies through convolutional operations and detecting features at various levels of abstraction, making them pivotal in \ac{CV} tasks. Transformers, on the other hand, leverage self-attention mechanisms to model long-range dependencies, excelling in handling sequential data such as text and time series. The integration of \acp{CNN} and Transformers enables more robust learning, particularly in tasks that require attention to both local and global data structures, such as object detection.} 

\begin{figure}[h]
    \centering
    \includegraphics[scale=0.9]{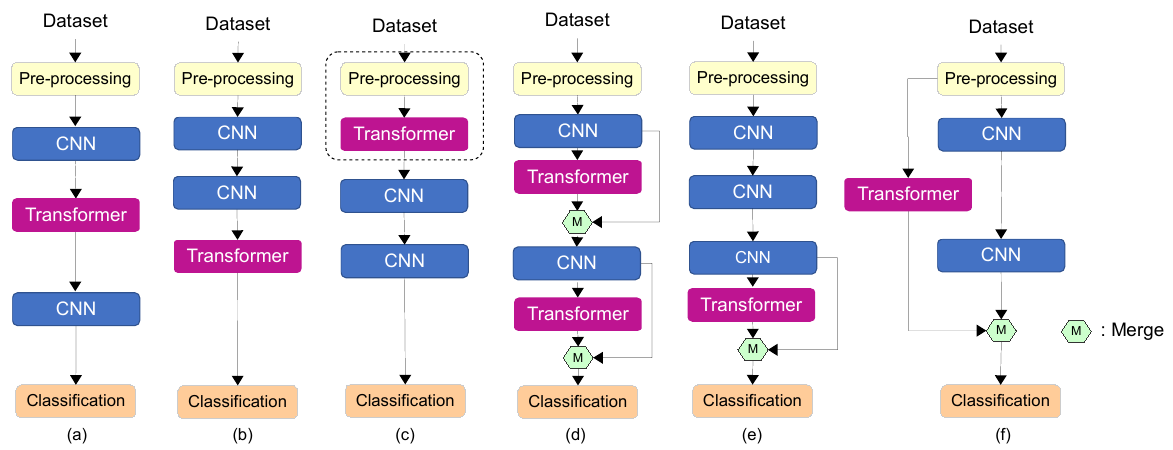}
    \caption{Various configurations of CNN-Transformer mechanisms include: (a) CNN and Transformer alternated \cite{duzgun2024network}; (b) A sequence of CNN followed by a single Transformer \cite{ullah2023transcnn, yao2023cnn, zhang2022efficient, chai2023ctsf, zeng2023iflv, bai2023hybrid, zhang2024efficient}; (c) Transformer followed by sequence of CNN  \cite{wang2021ddostc,ullah2024ids,yin2021intrusion,luo2022hierarchical}; (d) One-by-one CNN-Transformer feature fusion \cite{li2021dct}; (e) A sequence of CNN followed by the last CNN and a single Transformer feature fusion \cite{jiang2024bbo,wang2022microcontroller}; (f) Sequence of CNN and Transformer feature fusion \cite{li2022tcmal}. }
    \label{fig:cnnTrans}
\end{figure}

{The \ac{SOTA} configurations of CNN-Transformers used in the realm of \ac{IDS} are illustrated in Figure \ref{fig:cnnTrans}. Particularly, the scheme \cite{duzgun2024network} alternates between \ac{CNN} for spatial data processing and the Transformer for character-level text interpretation, following the configuration of Figure \ref{fig:cnnTrans} (a), enhancing anomaly detection through deep contextual understanding and efficient feature extraction. } {The configuration depicted in Figure \ref{fig:cnnTrans} (b) has been explored in numerous schemes.} For instance, feature subsets covering multiple spaces are developed in \cite{yao2023cnn} using \ac{CNN} to enhance the spatial distribution of samples. Subsequently, the Transformer is employed to establish connections between features and to identify essential attributes, including the temporal and detailed aspects of the features, culminating in the successful detection of intrusion activities.  Similarly, the work in \cite{he2022network} presents a hybrid neural network model aiming to improve feature extraction and detection effectiveness in traffic analysis. It combines dense \acp{CNN} and Transformers for feature fusion and time sequence extraction. Results on the CIC-IDS2018 dataset show 98\% accuracy, surpassing existing models in performance metrics. Moving on, the study referenced in \cite{lopes2023network} employs 1D \ac{CNN} and time series Transformer architectures to address the challenges posed by \acp{RNN} related to computational complexity and detection performance, particularly due to issues of information loss. The performance assessment on the CICDDoS2019 and CSE-CIC-IDS2018 datasets showed outstanding results, with the evaluation metrics mostly falling between 98.07\% and 99.99\%.

{All the methods in \cite{wang2021ddostc, ullah2024ids, yin2021intrusion, luo2022hierarchical} are Transformer-based \ac{IDS} approaches where the Transformer precedes the \ac{CNN} block, as depicted in the configuration of Figure \ref{fig:cnnTrans} (c). In the DDosTC model \cite{wang2021ddostc}, the Transformer, followed by a sequence of \ac{CNN}, work jointly to enhance \ac{DDoS} attack detection. The Transformer processes sequential data, focusing on contextual dependencies, while the \ac{CNN} extracts spatial features, optimizing for pattern recognition in network traffic. This synergy improves accuracy and robustness, outperforming models that rely solely on one type of network architecture. However, The DDosTC model may have limited generalizability due to testing on a single dataset and potential inefficiencies in real-time application scenarios with varied attack vectors. Similarly, In the IDS-INT model \cite{ullah2024ids}, Transformers first extract detailed feature representations from network interactions, focusing on semantic relationships within the data. After feature extraction, the \ac{SMOTE} is applied to balance the dataset by synthesizing new samples for minority classes. Subsequently, the \ac{CNN} processes these balanced features to deeply analyze and extract spatial patterns. This sequential processing significantly enhances the model's ability to detect a wide range of network intrusions effectively. However, the method's high computational demand could limit its applicability in resource-constrained environments. Yin et al. \cite{yin2021intrusion} proposed a mixed attention mechanism that enhances feature extraction by prioritizing influential attributes, while the \ac{CNN} handles the spatial relationships between these features. This combination significantly boosts the model's ability to discern subtle anomalies. Nevertheless, the capsule network's dynamic routing mechanism is slower than traditional networks, requiring efficiency improvements. }

The \ac{CNN} and Transformer in the method proposed by Li et al. \cite{li2021dct} follow the configuration shown in Figure \ref{fig:cnnTrans} (d). This method enhances anomaly detection by capturing multi-scale features and optimizing sequential dependencies. Skip connections improve training stability and feature utilization, further boosting the model's accuracy and robustness in time series data. The configuration shown in Figure \ref{fig:cnnTrans} (e) has been adopted by \cite{jiang2024bbo,wang2022microcontroller}, where both employed \ac{CNN} combined with \ac{MHA} Transformer to enhance feature extraction by capturing both local and global dependencies in the data. Skip connections facilitate better gradient flow, preventing vanishing gradients and enabling deeper network training. This integration leads to more accurate and robust detection of anomalies or network intrusions in complex datasets. However, in \cite{jiang2024bbo}, the hierarchical structure and deep architecture may result in increased computational complexity, potentially limiting real-time application capabilities, while in \cite{wang2022microcontroller}, the reliance on extensive network depth and \ac{MHA} mechanisms can lead to overfitting, particularly with imbalanced or insufficiently diverse datasets. The configuration depicted in Figure \ref{fig:cnnTrans} (f), where a sequence of \ac{CNN} layers is fused with Transformer features, has already been adopted by Li et al. in \cite{li2022tcmal}. The fusion of \ac{CNN} and Transformer encoder captures both spatial and contextual features, significantly enhancing the precision and robustness of encrypted malicious traffic classification. However, the model's complexity and resource-intensive pre-training increase computational demands, potentially limiting scalability and hindering deployment in real-time environments.

Researchers have proposed numerous \ac{SOTA} methods that combine \ac{CNN} with \ac{LSTM} to form hybrid schemes like CNN-LSTM, applied in various research fields including \ac{IDS} \cite{gueriani2024enhancing}, among others. These combinations have proven effective in enhancing overall accuracy. However, to handle long dependencies, researchers have increasingly either added Transformers or replaced \acp{LSTM} with Transformers. Wang et al. \cite{wang2023res} suggest a model for detecting intrusions by integrating the pre-trained ResNet model, \ac{MHA}-based Transformer, and bidirectional \ac{LSTM} models into a single framework that captures both spatial and temporal characteristics of network traffic, leveraging their robust capabilities in learning data representations. This approach significantly reduces the time required to process complex, high-dimensional intrusion data through \ac{DL} techniques. The objective is to address challenges in \acp{IDS}, such as inadequate feature extraction and imprecise classification with complex, nonlinear, and high-dimensional data. To ensure a balanced dataset, the \ac{SMOTE} has been utilized for preprocessing.

The developed Transformer-based model, referred to as XTM \cite{baul2023xtm}, aims to accurately detect and locate data breaches in real-time settings. This model, integrating Transformer and \ac{LSTM} technologies, marks the first attempt to evaluate the effectiveness of Transformer models in the smart grid research field, achieving excellent detection accuracy for the IEEE-14 bus system.

In their research, Ding et al. \cite{ding2024mf} proposed a method for securing Internet traffic, applicable to both IPv4 and IPv6. This method utilizes a multi-frequency \ac{LSTM}  alongside a multi-frequency Transformer module, each comprising layers designed to process high-frequency and low-frequency data. This approach enables the detection of temporal details as well as the frequency of attacks by analyzing Internet traffic patterns as a combination of sequential data that occurs at different frequencies.  The proposed work in \cite{fournier2021improving} introduces a method to enhance the analysis of kernel traces, which are sequences of low-level events, by incorporating event arguments. This approach involves learning a representation of event names and their arguments using embedding and encoding techniques. Evaluation through ablation studies on call-related, process-related, and time-related arguments demonstrates its effectiveness. Experiments on web request and server datasets show performance improvements of up to 11.3\% on unsupervised language modeling tasks using \ac{LSTM} and Transformer networks. These tasks aid in anomaly detection, neural network pre-training, and contextual event representation extraction.

\subsection{\ac{ViT}-based methods}

\begin{figure}[h!]
    \centering
    \includegraphics[scale=0.8]{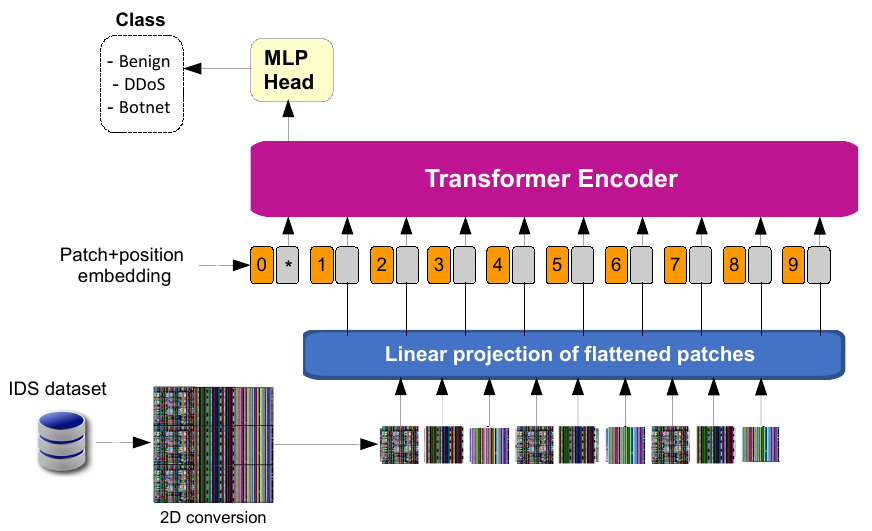}
    \caption{General structure of \ac{ViT}{-based IDS}, where an asterisk (*) indicates an additional learnable class embedding. The \acf{MLP} is responsible for feature transformation, global context fusion, and classification or regression.}
    \label{fig:ViT}
\end{figure}

\Acp{ViT} and \acp{CNN} offer distinct approaches for analyzing visual data. \Acp{ViT}  utilize the Transformer architecture to process images as sequences of patches, applying self-Attention mechanisms to understand global relationships within the image. This method contrasts with \acp{CNN}, which analyze images using convolutional filters that focus on local features and incrementally expand their understanding to more complex patterns. In other words,  \Acp{ViT} model relationships across the entire image using Attention, enabling dynamic focus on pertinent areas, whereas \acp{CNN} build a hierarchical understanding of local features through successive layers. Consequently, \Acp{ViT} might perform better in scenarios requiring a comprehend of the global context but usually need more data and computational power for training. In contrast, \acp{CNN} excel at efficiently learning spatial hierarchies but may not capture long-distance dependencies as effectively as \Acp{ViT}. Figure \ref{fig:ViT} illustrate a basic structure of \ac{ViT} Transformers. For example, \cite{li2022mfvt}  proposed architecture combines a feature fusion network with a \ac{ViT}, enhancing the overall \ac{DL} model's ability to handle imbalanced datasets and reducing the amount of sample data required for training. The authors in \cite{ho2022network} suggest an \ac{IDS} approach, wherein they utilize image conversion from network data flow, by mapping flow data to RGB values, to generate an RGB image. This image is then analyzed using the \ac{DT} algorithm to pinpoint significant features. Additionally, a \ac{ViT} classifier is employed to categorize the resulting image. Similarly, in their work, Agrafiotis et al. \cite{agrafiotis2022image} transformed traffic data contained in \ac{PCAP} files into grayscale images. Following this conversion, they utilized \ac{ViT} techniques for malware classification, subsequently evaluating its performance in comparison to that of \ac{CNN}. 

The aim of the work in \cite{zhu2023distributed} is to convert intrusion detection into a multiclass classification task for identifying intrusion events. In this research, the authors introduce a method that utilizes image encoding coupled with the  \ac{SwinT} model from computer vision for pattern recognition. Specifically, timing signals gathered are converted into 2-D images. This encoding strengthens the correlation and time dependency among sampling points, while the design of \ac{SwinT} incorporates window and shifted window techniques for multiscale feature extraction. Additionally, they employed the focus loss function to mitigate the impact of class imbalance in real-world scenarios.

Given the constant emergence of new vulnerabilities and unknown attack types, typically only a few samples of these attacks are available for analysis, which current detection methods in real systems struggle to handle. To address this issue, the authors in \cite{du2023few} suggest a few-shot class-incremental learning approach. This method allows for continuous learning of new attack classes using a minimal number of samples. It employs a \ac{ViT} as a self-supervised feature extractor and a dual-session branch classifier learning module. This module includes two phases: the base and the new session branch classifier learning. These phases are designed to adapt the parameters of the projection layer for different sessions and implement a fusion strategy to enhance the model's training and inference capabilities.

De Rose et al. in \cite{de2024vincent} proposed a VINCENT method  utilizes a \ac{ViT} as the teacher model in its knowledge distillation setup to detect and classify cyber threats. It incorporates an imagery representation of cyber-data, where the \ac{ViT} processes these images to generate attention-based features that are significant for classifying different types of threats. This methodology allows VINCENT to effectively recognize and categorize various cyber threats, achieving a notable performance with the best overall accuracy of 87.2\%. However, the batch-based learning approach which does not account for the evolving nature of cyber threats, potentially limiting its effectiveness over time as new threat patterns emerge. Likewise, the DE-VIT scheme proposed in \cite{he2024network} leverages \ac{ViT} capabilities for image classification techniques to tackle intrusion detection by transforming cyber threat data into a visual format. This method uses deformable attention mechanisms to focus selectively on parts of the data, rather than processing entire feature sets indiscriminately, thereby improving detection accuracy, especially in varied and complex datasets. The best result was achieved with a 99.5\% accuracy on the CIC IDS2017 dataset. However, a notable limitation of DE-VIT is that its performance decreases rapidly in data categories with fewer samples, highlighting an issue with handling imbalanced datasets. Unlike previous approaches, Yang et al. \cite{yang2024convitml} proposed the ConViTML approach, which integrates \ac{CNN} and \ac{ViT} to capture both local and global features in network traffic data. The \ac{CNN} efficiently extracts low-dimensional, shallow features from individual packets, while the \ac{ViT} captures the structural and sequential relationships between these packets. This fusion significantly enhances the model's capability to classify encrypted malicious traffic with high precision. The best performance was achieved an accuracy of 99.75\%. Even so, the model's complexity and reliance on extensive training data may limit its scalability and real-time deployment efficiency.

\subsection{GAN-Transformer-based methods}

\begin{figure}[h!]
    \centering
    \includegraphics[scale= 0.85]{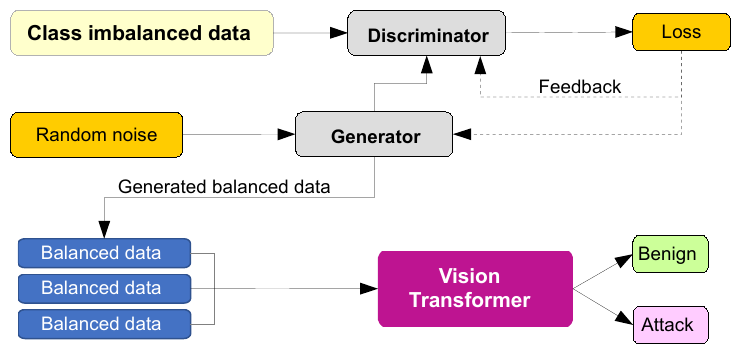}
    \caption{Illustration of the GAN and Transformer-based framework for anomaly detection when dataset is imbalanced.}
    \label{fig:TransGAN}
\end{figure}

The GAN-Transformer is a hybrid model that integrates \acp{GAN} with Transformer architectures to improve generative tasks across various domains. This model is particularly effective in fields where data can be transformed into text or image, allowing for the application of established \ac{NLP} techniques, as well as in image generation. By combining these two powerful technologies, the \ac{GAN}-Transformer leverages the strengths of \acp{GAN} in generating high-fidelity outputs with the Transformer's ability to handle complex data dependencies, thereby enhancing the model's overall performance in generating sophisticated and contextually relevant outputs. Figure \ref{fig:TransGAN} illustrates a general concept of utilizing GAN and \ac{ViT} to improve threat detection in imbalanced network flow data conditions. 

Integrating \acp{ViT} with \acp{GAN} involves adapting the traditional \ac{GAN} framework to leverage the Transformer architecture for either the generator, the discriminator, or both. While the core objective function of \acp{GAN} remains the same, the architecture of the generator and discriminator changes to incorporate Transformers. The core objective function of \acp{GAN} is  $\min_G \max_D V(D, G)$, where the value function $V(D, G)$  is given by:
\begin{equation}
 V(D, G) = \mathbb{E}_{x \sim p_{\text{data}}(x)}[\log D(x)] + \mathbb{E}_{z \sim p_z(z)}[\log (1 - D(G(z)))]    
\end{equation}

\noindent Where \( p_{\text{data}}(x) \) is the probability distribution of the real data. \( p_z(z) \) is the probability distribution of the input noise (e.g., a Gaussian distribution) fed into the generator. \( G(z) \) is the generator function that maps the noise \( z \) to the data space. \( D(x) \) is the discriminator function that outputs the probability that the input \( x \) is from the real data distribution. When \acp{ViT} are used, the functions $ G$ and $D$ are replaced with Transformer-based models:
\begin{itemize}
    \item \textbf{\ac{ViT} Generator \( G \):} This Transformer-based generator \( G \) takes a latent vector \( z \) and produces an image or feature map. The output of \( G(z) \) is typically processed by a series of Transformer blocks that generate high-quality images or features.
    \item \textbf{\ac{ViT} Discriminator \( D \):} This Transformer-based discriminator \( D \) takes an image or feature map \( x \) and outputs a probability that the input is from the real data distribution. 

\end{itemize}

\noindent Let \( G_{\text{ViT}} \) denote the \ac{ViT}-based generator and \( D_{\text{ViT}} \) denote the \ac{ViT}-based discriminator. The \ac{GAN} objective function with \acp{ViT} can be expressed as $\min_{G_{\text{ViT}}} \max_{D_{\text{ViT}}} V(D_{\text{ViT}}, G_{\text{ViT}}) $, where

\begin{equation}
V(D_{\text{ViT}}, G_{\text{ViT}}) = \mathbb{E}_{x \sim p_{\text{data}}(x)}[\log D_{\text{ViT}}(x)] + \mathbb{E}_{z \sim p_z(z)}[\log (1 - D_{\text{ViT}}(G_{\text{ViT}}(z)))]
\end{equation}

Although numerous researchers have introduced various \ac{GAN}-based methodologies to address the time series anomaly detection issue, challenges such as model collapse, limited generalization, and low accuracy persist. In this paper \cite{li2021dct}, the authors introduce a dilated convolutional Transformer-based \ac{GAN} aimed at increasing model accuracy and enhancing its generalization capabilities. The method employs multiple generators and a single discriminator to mitigate the problem of mode collapse. Each generator features a dilated \ac{CNN} paired with a Transformer block, which consists of \ac{MHA}, designed to capture both fine-grained and coarse-grained time series data, thereby boosting the model's generalization ability. Additionally, a weight-based mechanism is utilized to maintain equilibrium among the generators.

The classification performance for intrusion detection suffers due to imbalanced training data and restricted feature extraction. The researchers in \cite{li2023pre}, introduces a novel method combining a \ac{CGAN}  with \ac{BERT}, a pre-trained language model, for multi-class intrusion detection. This technique addresses the imbalance in attack data by leveraging \ac{CGAN}  to augment minority class samples. Additionally, \ac{BERT}, known for its robust feature extraction capabilities, is integrated into the \ac{CGAN}  discriminator to enhance the relationship between input and output, thereby improving detection accuracy through adversarial training. Similarly, \cite{wang2022research} integrates \ac{CGAN}  with \ac{ViT} to enhance the accuracy of network traffic data detection, particularly when confronted with imbalanced network flow data conditions. Only the encoder segment of the \ac{ViT} model is utilized.

\subsection{Other Transformer-based methods}
\Ac{FL} is a decentralized DL approach where model training occurs locally on devices or servers holding data. Instead of sending data to a central server, only model updates or gradients are shared. These updates are aggregated to improve the global model, which is then redistributed. This process preserves data privacy, as raw data remains on local devices, mitigating privacy risks associated with centralized data storage. \ac{FL} also enables collaborative model training across distributed environments, benefiting from diverse data sources without compromising individual privacy. It finds applications in healthcare, finance, and other sectors where data privacy is paramount \cite{himeur2023federated}. Employing Transformers together with \ac{FL} has been investigated by researchers. For example, in \cite{zhou2024network}, a \ac{NIDS} method employing \ac{FL} and an enhanced Transformer model, which includes \ac{MHA}, addresses issues of prolonged detection time and low accuracy. Data augmentation and local model analysis enhance detection, with final predictions aggregated using a Softmax classifier. Similarly, the work  \cite{abdel2021federated}  presents FED-IDS, a FL-based \ac{IDS} that offloads learning to vehicular edge nodes. It uses a context-aware Transformer network with \ac{MHA} to capture the spatial-temporal patterns of abnormal and normal vehicular data,   and blockchain-managed federated training for secure, distributed, and reliable attack detection. Similarly, in \cite{sun2022hierarchical}, the authors claim that existing \ac{IDS} models have low performance and are typically trained on cloud servers, which jeopardizes user privacy and increases detection delay. To address these issues, they present a Transformer-based model to enhance \ac{IDS} performance. Additionally, it integrates 5G technology into smart grid systems and proposes HFed-IDS, a hierarchical \ac{FL} system, to collaboratively train the proposed Transformer-based \ac{IDS} model and protect user privacy in core networks.

An n-gram is a contiguous sequence of n items from a given sample of text. It is used instead of a single word because it provides valuable context information. Han et al. \cite{han2023network}, proposed a novel intrusion detection model called GTID, which leverages n-gram frequency and a time-aware Transformer. GTID hierarchically learns traffic features from both packet-level and session-level data, minimizing information loss. It processes packet headers and payloads differently to extract packet-level features effectively, using n-gram frequency to capture payload context. For session-level features, GTID employs a time-aware Transformer with \ac{MHA}, considering time intervals between packets to learn temporal session features for accurate intrusion detection.

\section{LLM-based methods}
\label{sec4}

The term \ac{LLM} is used to differentiate language models by the size of their parameters, specifically those considered large-sized pre-trained models. However, the academic community has not reached a formal agreement on the minimum parameter size required for a model to be classified as an \ac{LLM}, as the model's capacity is closely related to the size of the training data and the overall computational resources available \cite{xu2024large}. The categories of \acp{LLM} can be divided into three types:

\begin{itemize}
    \item \textbf{Encoder-only \acp{LLM}:} are a type of \acp{LLM} that primarily utilize the encoder component of the Transformer architecture and focus on understanding and generating representations of the input data. These models excel at tasks that involve understanding and classifying text, such as sentiment analysis, named entity recognition, and text classification. \Ac{MLM} is a technique used primarily in encoder-only models, like \ac{BERT}, to train the model on understanding context by predicting missing words in a sentence. In \Ac{MLM}, some tokens in the input sequence are randomly replaced with a special "[MASK]" token, and the model is trained to predict the original tokens at these masked positions. The goal is to maximize the likelihood of predicting the original tokens given the masked sequence. Given an input sequence \( x = (x_1, x_2, \ldots, x_n) \), where some tokens are masked (replaced with `[MASK]`), the model's task is to predict the original tokens at the masked positions. Let \( M \) be the set of positions that are masked. For each masked position \( i \in M \), the model outputs a probability distribution \( P(x_i \mid x_{1}, x_{2}, \ldots, x_{n}) \) over the vocabulary. The objective is to maximize the likelihood of the original tokens \( x_i \) at the masked positions. The \ac{MLM} loss function is defined as:
    \begin{equation}
            \mathcal{L}_{\text{MLM}} = -\sum_{i \in M} \log P(x_i \mid x_{1}, x_{2}, \ldots, x_{n})
    \end{equation}
Where, \( P(x_i \mid x_{1}, x_{2}, \ldots, x_{n}) \) is the probability assigned by the model to the actual token \( x_i \) given the entire sequence \( x \), where the model has seen the context around the masked tokens. Cross-Entropy Loss aims to measures how well the model's predicted probability distribution matches the actual token at the masked positions.

    \item \textbf{Decoder-only \acp{LLM}:} are a type of language model that primarily utilize the decoder component of the Transformer architecture. Unlike encoder-only or encoder-decoder models, decoder-only models focus on generating text by predicting the next token in a sequence based on the previous tokens. These models are particularly well-suited for tasks involving text generation, such as language modeling, dialogue systems, and creative writing.
    
    In decoder-only \ac{LLM}, autoregressive modeling is employed to predict the next token in a sequence based on the preceding context. This approach is used by models like \ac{GPT}. The primary loss function used for autoregressive modeling in decoder-only \acp{LLM} is the cross-entropy loss. This loss function is designed to maximize the likelihood of the model correctly predicting each token in the sequence. For an autoregressive model, given a sequence of tokens \( x = (x_1, x_2, \ldots, x_n) \), the model is trained to predict each token \( x_i \) given the previous tokens \( (x_{1}, x_{2}, \ldots, x_{i-1}) \). The cross-entropy loss function is defined as:
\begin{equation}
\label{eqLCE}
    \mathcal{L}_{\text{CE}} = -\sum_{i=1}^{n} \log P(x_i \mid x_{1}, x_{2}, \ldots, x_{i-1})
\end{equation}

Where \( P(x_i \mid x_{1}, x_{2}, \ldots, x_{i-1}) \) is the probability assigned by the model to the actual token \( x_i \) given the previous tokens \( (x_{1}, x_{2}, \ldots, x_{i-1}) \). Cross-entropy loss  measures the performance of the model by comparing the predicted probability distribution to the actual token distribution. The probability of generating the entire sequence \( x \) in an autoregressive model is given by:

\begin{equation}
 P(x) = \prod_{i=1}^{n} P(x_i \mid x_{1}, x_{2}, \ldots, x_{i-1}) 
\end{equation}

The training objective is to maximize the likelihood of the sequence \( x \) being generated, which is equivalent to minimizing the negative log-likelihood described in Equation \ref{eqLCE}.

\item \textbf{Encoder-decoder \acp{LLM}:} also known as \ac{seq2seq} models, utilize both the encoder and decoder components of the Transformer architecture. These models are designed to transform a sequence of input data into a sequence of output data, making them well-suited for tasks where an input sequence needs to be mapped to an output sequence.
    
In encoder-decoder \acp{LLM}, such as BERT2GPT or T5, autoregressive modeling is used in the decoding phase to generate sequences based on the encoded input. The model consists of two main components: (i) \textbf{An encoder}, which processes the input sequence and generates context or embeddings. (ii) A \textbf{decoder}, generates the output sequence autoregressively based on the encoder's output and previously generated tokens. For an input sequence \( x = (x_1, x_2, \ldots, x_m) \) and a target output sequence \( y = (y_1, y_2, \ldots, y_n) \), the decoder is trained to predict each token \( y_i \) given the previously generated tokens \( (y_{1}, y_{2}, \ldots, y_{i-1}) \) and the encoded representation from the encoder. The cross-entropy loss function for autoregressive modeling in encoder-decoder \acp{LLM} is defined as:

\begin{equation}
\label{eqLCED}
 \mathcal{L}_{\text{CE}} = -\sum_{i=1}^{n} \log P(y_i \mid y_{1}, y_{2}, \ldots, y_{i-1}, \text{Encoder}(x))   
\end{equation}

Where \( P(y_i \mid y_{1}, y_{2}, \ldots, y_{i-1}, \text{Encoder}(x)) \) is the probability assigned by the decoder to the actual token \( y_i \) given the previous tokens \( (y_{1}, y_{2}, \ldots, y_{i-1}) \) and the encoder’s output for the input sequence \( x \).  The probability of generating the target sequence \( y \) given the input sequence \( x \) in an autoregressive manner is given by:

\begin{equation}
 P(y \mid x) = \prod_{i=1}^{n} P(y_i \mid y_{1}, y_{2}, \ldots, y_{i-1}, \text{Encoder}(x))
\end{equation}

The training objective is to maximize the likelihood of the target sequence \( y \) given the input sequence \( x \), which is equivalent to minimizing the negative log-likelihood described in Equation \ref{eqLCED}.
\end{itemize}

\begin{figure}[h!]
    \centering
    \includegraphics[scale=0.8]{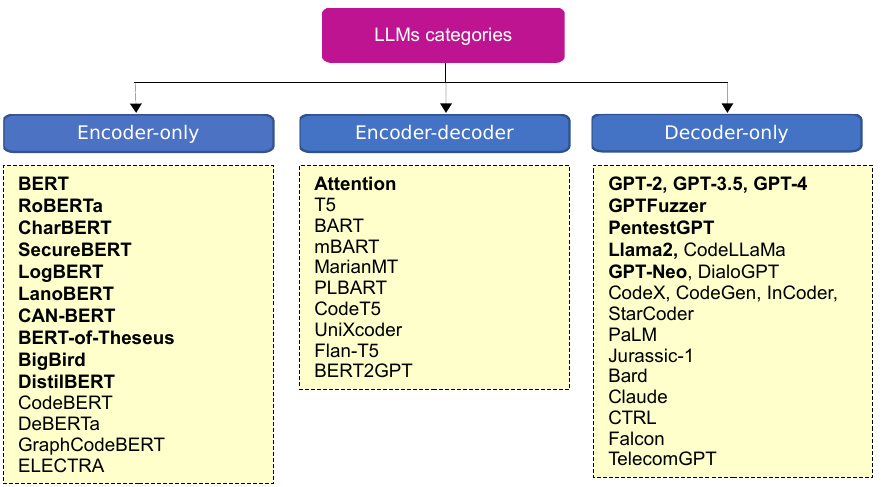}
    \caption{A taxonomy categorizing \acp{LLM} into three distinct groups. {The \acp{LLM} marked in bold are those already utilized for \ac{IDS} and reviewed in this survey.} }
    \label{fig:llmCat}
\end{figure}

\noindent Figure \ref{fig:llmCat} illustrates a taxonomy of the three categories, summarizing the most well-known LLM models for each category.

\Ac{LLM} can enhance \ac{NLP}-based computer security solutions, enabling more effective detection and classification of malicious content, phishing attempts, and malware. Additionally, \ac{GPT} \ac{LLM} can contribute to the development of conversational agents for security operations, facilitating quicker responses to security incidents and providing valuable insights into emerging threats. Table \ref{table:ResFind} summarizes various research findings on \acp{LLM}-based IDS schemes, including the dedicated tasks, comparison methods, datasets used, obtained results, and improvements achieved.

\subsection{Encoder-only-based methods}

\noindent \Ac{BERT} is a \ac{LLM} developed by Google. It utilizes the Transformer architecture, which comprises self-Attention mechanisms to capture contextual relationships between words bidirectionally. The model consists of multiple layers of Transformers, with each layer containing self-Attention and feed-forward neural networks.  \ac{BERT} employs a pre-training and fine-tuning approach. During pre-training, it learns contextualized representations of words by predicting masked words within a sentence and predicting sentence-level relationships in a language modeling objective. Fine-tuning involves further training on downstream tasks, such as sentiment analysis or named entity recognition, by adjusting parameters based on task-specific data.  \noindent While \ac{GPT} is unidirectional and pre-trained with autoregressive language modeling. \ac{BERT}'s bidirectional nature and ability to capture rich contextual information have led to significant improvements in various \ac{NLP}  tasks.

\ac{BERT} offers benefits in intrusion detection by leveraging its contextual understanding of language. It can analyze network logs, system alerts, and other textual data to identify anomalous behavior or potential security threats more accurately. By capturing the nuances of language, \ac{BERT} can discern subtle patterns indicative of intrusions or attacks, enhancing the detection capabilities of intrusion detection systems. Additionally, its ability to handle unstructured data makes it effective for processing diverse sources of information commonly encountered in cyber-security applications.

\begin{figure}[h!]
    \centering
    \includegraphics[scale=0.75]{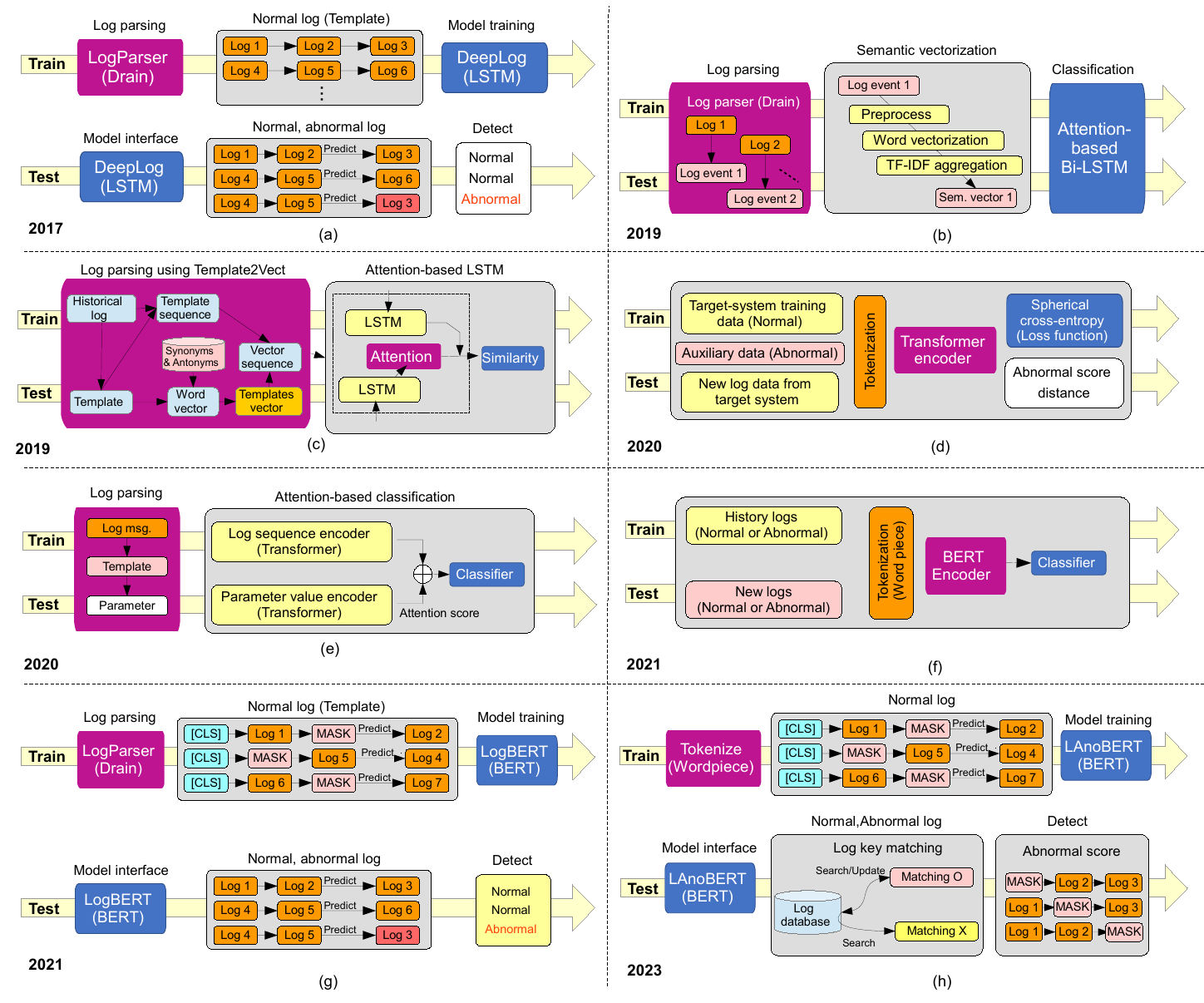}
    \caption{Benchmarking BERT-based LLM models for log anomaly detection: (a) \textbf{DeepLog}—focused on log anomaly detection; (b) \textbf{LogRobust}—employs robust statistical techniques; (c) \textbf{LogAnomaly}—utilizes a statistical analysis framework; (d) \textbf{LogSy}—leverages \ac{DL} for log generation; (e) \textbf{HitAnomaly}—detects anomalies in web server access logs using statistical methods; (f) \textbf{NeuralLog}—for comprehensive log anomaly detection and analysis; (g) \textbf{LogBERT}—targets log sequence classification and anomaly detection; (h) \textbf{LAnoBERT}—specialized for log anomaly detection tasks   \cite{lee2023lanobert}.}
    \label{fig:LAnoBERT}
\end{figure}

For example, \cite{wang2024lightweight}  introduces BT-TPF framework, an \ac{IoT} intrusion detection model using knowledge distillation. It employs a Siamese network for feature reduction and a BERT-of-Theseus Transformer, which involves compressing the BERT model by replacing its modules, as a teacher model, achieving high accuracy with only 788 parameters, a 90\% reduction. Similarly, the authors in \cite{nguyen2023method} explore leveraging information from a sequence of network flows to enhance the domain adaptation capability of the \ac{NIDS}. They propose a framework  that utilizes \ac{BERT} for feature extraction and \ac{MLP} for classification. Moreover, named entity recognition  is vital text  in structuring complex cyber-threat intelligence for cybersecurity. However, existing research has focused largely on English CTIs, with poor performance in Chinese. To address this, RoBERTa-wwm \ac{LLM} is proposed in \cite{zhen2023chinese}, utilizing Chinese pre-trained language models to effectively handle English-Chinese word mixing in cyber threat intelligence, such as intrusions. Moving on, Aghaei et al. \cite{aghaei2023automated}  present their innovative predictive model and tool based on SecureBERT \ac{LLM} model  that can generate priority recommendation reports for potential cybersecurity threats and predict their impact.

The system log produced by a computer system comprises extensive data gathered simultaneously, serving as the foundational data for identifying errors, intrusions, and abnormal behaviors. Detecting anomalies in system logs aims to swiftly identify irregularities with minimal human intervention, a critical challenge in the industry. An approach previous suggested typically involved converting log data into a standardized template using a parser before anomaly detection algorithms could be applied. For instance, LogBERT \cite{guo2021logbert}, a BERT-based anomaly detection framework, refines log sequences using a drain parser. It trains solely on normal log data through two tasks: masked log key prediction and volume of hypersphere minimization. The first task trains normal log patterns using masked language modeling, while the second task identifies the smallest sphere containing normal logs. During inference, top predicted log keys form a candidate set from a randomly masked normal log sequence. An observed log key not belonging to this set indicates an anomaly. However, LogBERT's random log key selection during masking limits its ability to fully consider the log sequence. Defining templates for specific events in advance could lead to loss of information within the log keys. In the study conducted by Lee et al. \cite{lee2023lanobert}, they introduced LAnoBERT, a parser-free method for system log anomaly detection that leverages the \ac{BERT} model, demonstrating strong \ac{NLP} capabilities. LAnoBERT learns the model through masked language modeling, a BERT-based pre-training method, and employs unsupervised learning-based anomaly detection using the masked language modeling loss function per log key during testing.  Experiments demonstrated that LAnoBERT not only outperforms unsupervised learning-based benchmark models in anomaly detection but also achieves comparable performance to supervised learning-based benchmark models. {Moving forward,  the paper in \cite{almodovar2024logfit}, proposes the LogFiT method, an advanced log anomaly detection model using a fine-tuned BERT-based language model. Unlike previous methods such as DeepLog and LogBERT, which rely on log templates for analysis, LogFiT operates directly on raw logs without requiring these templates, significantly improving anomaly detection accuracy through enhanced semantic analysis. The model leverages a self-supervised training regime, utilizing masked sentence prediction to deeply understand normal operational log data, making it highly effective in spotting deviations indicative of system threats. However, the method's limitation includes high computational demands and a dependency on extensive pre-training.}

Figure \ref{fig:LAnoBERT} depicts the proposed Transformers and LLM-based intrusion and anomaly detection, a modified version of BERT or Attention Transformers, serving as the foundation of an IDS for such models.  The figure highlights the essential components of Transformers and LLMs, providing a conceptual representation focused on the core components.  An \ac{LLM} model, such as BERT, can be substituted with one of the Transformer or \ac{LLM} models indicated in Figure \ref{fig:LAnoBERT} and paired with a decoder from Figure \ref{fig:llmCat}. This configuration yields many new encoder-decoder \acp{LLM}, developed in accordance with the BERT2GPT principle, specifically for \ac{IDS} applications. Additionally, merging model blocks from the same figure could produce a specialized encoder dedicated to \ac{IDS}.

\subsection{Decoder-only-based methods}

\begin{figure}[h!]
     \centering
     \includegraphics[scale=0.8]{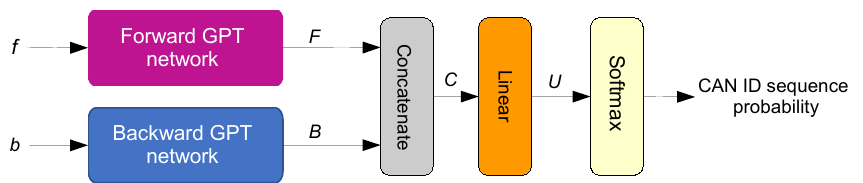}
     \caption{An example of employing two \ac{GPT} \ac{LLM}s to detect attacks. The process of detecting intrusions for the given CAN ID sequence \( x \) involves converting it into \( f \) and \( b \) sequences using the bidirectional GPT network structure. These sequences are then fed into the forward and backward GPT networks, respectively. The forward GPT network processes \( f \) and generates the \( F \) matrix, while the backward GPT network processes \( b \) to generate the \( B \) matrix. These \( F \) and \( B \) matrices are then combined to form a composite \( C \) matrix. Then, the top linear layer \( W \) takes \( C \) as its input and outputs a matrix with dimensions \( U \). Finally, the estimated probability for the CAN ID sequence \( x \) is obtained from the output of the softmax activation function  \cite{nam2021intrusion}.}
     \label{fig:bgpt}
 \end{figure}

The \ac{GPT} as \ac{LLM} is an AI model developed by OpenAI, adept at understanding and generating human-like text. It utilizes the Transformer architecture, employing self-Attention mechanisms to capture contextual dependencies in language. Trained on extensive text data, \ac{GPT} \ac{LLM} learns to generate coherent and contextually relevant responses across diverse tasks. Its role in cyber-security is multifaceted. \ac{GPT} \ac{LLM} can assist in threat intelligence by analyzing and summarizing large volumes of security reports and logs, aiding in the identification of potential threats. For example, the research in \cite{chen2023applying} presents an approach that combines the diamond model of intrusion analysis with \ac{DL} techniques to offer a holistic understanding of malware attacks. The authors  investigate the effectiveness of \ac{BERT} and \ac{GPT} 
 LLM models in generating threat intelligence reports (Answers) from a list of pre-defined queries (questions). The findings demonstrate that \ac{GPT}  outperformed \ac{BERT} in terms of performance. Another research example is investigated in \cite{nam2021intrusion} for \ac{CAN} bus protocol security shortage. The \ac{CAN} bus protocol is vulnerable to various attacks. Existing detection methods struggle to identify patterns when only a few attack IDs are present in a \ac{CAN} ID sequence ( Figure \ref{fig:can}). A proposed solution involves utilizing a \ac{GPT} model to learn normal \ac{CAN} ID sequences' patterns. This approach outperforms traditional \ac{LSTM}-based methods. By combining two \ac{GPT} networks bidirectionally, past and future \ac{CAN} IDs can be considered, improving intrusion detection accuracy. Training aims to minimize negative log-likelihood values, with intrusions identified when exceeding a set threshold. Figure \ref{fig:bgpt} illustrates the principle of employing two \ac{GPT} \ac{LLM}s to detect attacks in a \ac{CAN} bus.

Web fuzzing is a security testing technique that generates and sends random or mutated inputs to a web application to discover vulnerabilities. It uses coverage feedback from the application to refine its inputs, helping identify potential security flaws.  The work proposed in  \cite{liang2023generative} introduced GPTFuzzer, which employs an encoder-decoder architecture to produce effective payloads for \acp{WAF}, specifically targeting \ac{SQL} injection, \ac{XSS}, and \ac{RCE} attacks by generating fuzz test cases. This model undergoes \ac{RL} \cite{gueriani2023deep,kheddar2024reinforcement},  \ac{TL} and applies a \ac{KLD} \cite{kheddar2023deep} penalty to efficiently generate attack payloads and avoid local optimum issues. Similarly,  \cite{liu2020deepsqli} utilized an encoder-decoder architecture model to create \ac{SQL} injection detection test cases for web applications, translating user inputs into new test cases.

In contemporary risk management strategies, \ac{CTI} reporting is crucial.  With the increasing volume of \ac{CTI} reports, automated tools for report generation are becoming essential. The application of \acp{LLM} in network threat analysis can be divided into \ac{CTI} generation and \ac{CTI} analysis for decision-making. \ac{CTI} generation involves extracting \ac{CTI} from network security text information such as books, scientific research, technical reports.  Generating structured \ac{CTI} reports from unstructured information \cite{siracusano2023time}, and creating \ac{CTI} from network security entity graphs \cite{perrina2023agir}.  Moskal et al. \cite{moskal2023llms} explored using ChatGPT to assist or automate response decision-making for threat behaviors, demonstrating the potential of \ac{LLM} in handling simple network attack activities.

Penetration testing, a type of PROBE attack that causes intrusions, involves conducting controlled attacks on computer systems to assess their security, remains a key approach for organizations to enhance their defenses against intrusions and cyber threats. The general penetration testing process includes the following steps: (i) information gathering, (ii) payload construction, and (iii) vulnerability exploitation. In this context, Temara \cite{temara2023maximizing} utilized \ac{LLM}, specifically ChatGPT, for information gathering in penetration testing, including IP addresses, domain information, vendor technologies, SSL/TLS credentials, and other details of the target website. Similarly, Sai Charan et al. \cite{charan2023text} critically examined \acp{LLM} capability, such as Google's Bard and ChatGPT, to generate malicious payloads for penetration testing, with results indicating that ChatGPT can produce more targeted and complex payloads for attackers. Moreover, the researchers in  \cite{happe2023evaluating} developed an automated Linux privilege escalation guidance tool using \acp{LLM} including  GPT-4,  GPT-3.5-turbo, and  Llama2. Additionally, the automated penetration testing tool PentestGPT \cite{deng2023pentestgpt}, based on \acp{LLM}, outperformed GPT-3.5 and demonstrated excellent performance on a penetration testing benchmark with 13 scenarios and 182 subtasks by combining three self-interacting modules: inference, generation, and parsing modules.

The work recently proposed by Li et al. \cite{li2024dollm} introduced the DoLLM model, which enhances \ac{DDoS} detection using \ac{LLM} by structuring non-contextual network flows into sequential data that \acp{LLM} can process effectively. Specifically, the DoLLM method organizes flows into contextual sequences, converts them into tokens, and leverages the \ac{LLM}'s semantic space to enhance detection accuracy, focusing on the potent Llama2-7B model for its robust processing capabilities. The approach significantly advances the detection of complex \ac{DDoS} attacks, like carpet bombing, by effectively capturing and analyzing network flow data. The best result obtained by the DoLLM model is a high F1 score of 96.4\% in zero-shot scenarios, indicating robust detection capabilities. However, the DoLLM method suffers from high computational overhead, reliance on external processing, and challenges in real-time detection due to latency.

\begin{scriptsize}
\begin{longtable}{lm{2cm}m{5.5cm}m{2cm}m{1.5cm}m{2cm}m{1.5cm}}
\caption{Synopsis of certain advancements suggested in Transformer-based \ac{IDS}. The symbol ($+$) indicates improvement, while ($-$) denotes a decrease in performance or false classification. When many tests or comparisons are conducted, only the best result and/or maximum improvement are mentioned.} 
\label{table:ResFind} \\
\toprule
Ref. & Transformer   & Dedicated task and limitations&  Compared to & Dataset & Result (\%) & $\Delta$ PFP (\%) \\
\midrule
\endfirsthead

\multicolumn{7}{c}%
{{\bfseries \tablename\ \thetable{} -- Continued from previous page}} \\
\toprule
Ref. & Transformer   & Dedicated task and limitations &  Compared to & Dataset & Result (\%) & $\Delta$ PFP (\%) \\
\midrule
\endhead

\midrule \multicolumn{7}{r}{{Continued on next page}} \\ \bottomrule
\endfoot

\bottomrule
\endlastfoot

\hline \toprule
\cite{zhou2024network} & FL-MHA & Tackle prolonged detection times and enhance security and accuracy. The proposed method  needs to consider more intrusion types & Res-Tran BiLSTM & NSL-KDD  \newline UNSW-NB15  & Acc= 99.45 \newline Acc= 89.83& Acc= +3.08  \newline  Acc= +3.18 \\\hline

\cite{nguyen2022flow}  & BERT  &  It relies on representation of network flow sequences for classification. The distribution of flows within a sequence can be altered in smaller datasets. &  DT & CIDDS &  Acc= 99.4 & Competitive \newline (Internal) \\\hline

\cite{nam2021intrusion}  &  Bi-GPT  &  The IDS use sequences of CAN.  Detection of abnormal patterns in the messages' protocol field was not conducted. &  Uni-GPT & Elaborated & AUC= 99.8 & AUC= +0.3 \\\hline

\cite{tang2023intrusion}  &  ViT  &  Compare ViT to pre-trained CNN models for IDS applications. The computational resources and time required by different models can vary significantly.  &  H2O & EdgeIIoT &  Acc= 99.69 & Competitive \\\hline

    \cite{nguyen2023transformer} & Self-Attention & The method detects intrusions in vehicle systems. It is dedicated solely to application layer attacks with long training time. & CNN-LSTM &  Car Hacking & F1= 99.47 &F1= +0.24 \\\hline

\cite{alkhatib2022can}  & CAN-BERT  & Identify cyber threats on the CAN network. Requires substantial computational demands and memory usage. &  LSTM-AE & In-vehicle &  F1= 81--99 & F1 $\approx$ +10 (Spark)\\\hline

\cite{wang2023res}  &  MHA  &  Enhance the feature extraction capability of the IDS model for IoT. It valid only for supervised learning. &  ResNet18-BiLSTM &  MQTTset & Acc= 99.56 & Acc= +10  \\\hline

\cite{nwafor2022canbert}  & CANBERT  &  IDS for automotive network security. Requires high-resource environments and faces challenges in generalizing the model to different domains.  &  GIDS & OTIDS &  Acc= 100 & Slightly better  \\\hline

\cite{li2021dct}  &  GAN-Transformer  &  The method detects anomalies in time series data. It  has a time-consuming issue. &  TadGAN &  NAB &  F1= 70.7  & F1= +5.4 \\\hline

\cite{abdel2021federated}  &  MHA  &  The method classifies various types of attacks on vehicular traffic flows. It does not address the interpretability of classification decisions in IDS.  &  DeepFed & Car Hacking &  Acc= 97.82 & Acc= +1.69 \\\hline

\cite{karlsen2024large} &  { GPT-Neo} & {Differentiating between normal and abnormal behaviors using an unsupervised learning method. However, the method reliance on predefined rules for log parsing, and computational costs.} &{GPT-2 } & {Web server log}& {F1= 99.25} & {F1=+2.92}\\ \hline

\cite{wang2023robust}  &  Self-Attention  &  Unsupervised NIDS through self-supervised masked context reconstruction. {Different datasets' sensitivities complicate tuning for optimal loss components.} &  NeuTraL AD & KDD &  Acc= 99.98 & Acc= +0.25 \\\hline

\cite{hsiao2023detection} & MHA & The method detects both SQL injection and XSS threats. It is specifically tailored for application layer attacks only. & Multi-model
CNN-GRU  & Mixed & F1= 99.68 \newline (FPR= 0.22) & F1=+0.08  \newline (FPR= -0.32) \\\hline

\cite{barut2022r1dit} & NHA & The method  classifies malware traffic while preserving privacy. It needs to consider more intrusion types rather just enduring attacks. & feature-based methods &  CICIDS2017 & F1= 97.2 & F1= +19.2  \\\hline

\cite{dang2023using} & MHA & Detect intrusion in a dynamic environment. It has high computational complexity and long training times, and it cannot detect replay attacks.  & CatBoost &  CICIDS-2017 & Acc= 86.74 & Acc= +0.03 \\\hline

\cite{li2022tcmal}  &  MHA  & Enhance Transformer classification performance by adding positional information to IDS features. {Computationally intensive model, which may limit its practicality for real-time applications.} &  CNN-Transformer & UNSW-15 &  Acc= 87.50 & Acc= +0.5 \\\hline

\cite{han2023network}  &  MHA &  Intrusion detection using payload/session length, packet intervals, n-gram Transformer. {The packet feature extraction process is time-consuming.} &  LSTM & ISCX2012 &  F1= 99.37 & F1= +0.52  \\\hline

\cite{liu2023intrusion}  &  MHA & Designed for lengthy training, accurate for binary and multi-class intrusion. {The model's multi-class detection capability needs significant improvement.} & CNN-Transformer & NSL-KDD &  F1= 88.2 & F1= +2.1 \\\hline

\cite{zhang2022efficient}  &  MHA  &  Classify encrypted malicious data flow with CNN-Transformer techniques. Computationally intensive model limits practicality for real-time applications. &  TCMal-WT & STRA    & F1= 95.56 & F1= +9.7 \\\hline

\cite{yao2023cnn}  &  CNN-Transformer & The method employs \ac{XGBoost}. It  lacks for real communication environments. &  CNN-only, \newline Transformer-only,  \newline CNN-LSTM & KDDCup99   & Acc= 97.85  & Competitive with NDAE\\\hline

\cite{he2022network}  &  CNN-Transformer  &  Effectively extracting features and enhancing intrusion detection. {The model struggles with small training sample categories like SQL injection.}&  CNN-LSTM & CIC-IDS2018 &  Acc= 98.9 & Acc= +0.3 \\\hline

\cite{luo2022hierarchical}  &  MFVT   &   Method for detecting anomaly traffic. {The scheme restrict its applicability to other datasets or real-world conditions.} &  ViT &  IDS 2017 &  F1= 99.99 & Competitive\\\hline

\cite{zeng2023iflv}  &  ViT-LSTM-FCN  &  The method improves IDS performance in small to medium-sized datasets. {Lower detection performance for flooding attacks limits its effectiveness.}  &  HNN & AWID &  Acc= 99.97 & Acc= +0.08 \\\hline

\cite{ding2024mf}  &  MF-Transformer  &  Dual-frequency IDS for IPv4 and IPv6 traffic data. {The model relies heavily on accurate historical data of attack frequencies.} &  DAGMM \newline GRU & UNSW-NB15 \newline IPv6 data &  F1= 91.4 (IPv4) \newline F1= 99.94 (IPv6) & F1= +0.6 (IPv4) \newline F1= +0.19 (IPv6)\\\hline

\cite{ho2022network}  & ViT  &  The method converts network flow to images and employing ViT for thread classification. {Multi-class classification results are inaccurate due to dataset imbalance.} &  DBN-KELM &  CIC IDS2017 &  Acc= 98.5 & Acc= +8.09 \\\hline

\cite{du2023few}  &  ViT  &  The method is efficient for few-sample learning approach for NIDS. {Handling new vulnerabilities and unknown attack types remains unverified.} &  C-FSCIL & CIC-IDS2017 &  F1= 92.95 & F1= +1.05 \\\hline

\cite{li2023pre} &  GAN-BERT &  Improving detection of intrusions across multiple classes. {Limited to distinguish similar or highly concealed attacks.}&  Baselines & NF-ToN-IoT-V2 &  F1= 98.799 & F1= +13.779 \\\hline

\cite{lee2023lanobert}  &  LAnoBERT  &  Anomaly detection in system logs utilizing masked LLM. {Dependent on log parser compatibility rather than the logic of anomaly detection models} &  NeuralLog \newline LogSy & HDFS,  BGL \newline Thunderbird &  F1= 96.45, 90.83 \newline F1= 99.90 & Competitive \\\hline

\cite{wang2024lightweight}   &  BERT-of-Theseus  &  IDS for IoT using knowledge distillation. {Knowledge distillation might lose some performance nuances from the teacher model.} &  SVM-GAC \newline STFA-HDLID  & CIC-IDS2017 \newline TON\_IoT &  Acc= 99 &  Competitive\\\hline

\cite{zhen2023chinese}  &  RoBERTa-wwm  &  Named entity recognition cyber-thread detection. {Relies heavily on the quality and quantity of the labeled training data} &  BERT-RDCNN-CRF  & NER &  F1= 82.35 & F1= +3.53\\\hline

\cite{chen2023applying}  &  GPT  &  Threat
intelligence reporting. {The method  tested only on malware attacks.} &  BERT & Elaborated &  Acc.= 61 & Acc.= +17 \\\hline

\cite{wang2022network}  &  GAN-Transformer  &  Detection of anomalies in network traffic. {The GAN-Transformer combination creates a complex model architecture.} &  DT & CICIDS2017 &  Acc= 97.24 & Acc= +3.19 \\\hline

\cite{sanchez2024transfer}&  {BigBird} & {Integrating \acp{LLM} with system call analysis provides an  enhanced malware detection capabilities. However, it lacks external processing independence and adaptation for detecting zero-day attacks with benign data.} &{BERT, DistilBERT, GPT-2, Mistral, Longformer } & {MalwSpecSys}& {Acc= 86.67} & {Acc=+ 0.51}\\ 
\bottomrule
\end{longtable}
\begin{flushleft}
Abbreviations: Performance in fixed point (PFP), fully connected network (FCN)   
\end{flushleft}
\end{scriptsize}

\section{Environments and applications for Transformers- and LLM-based IDS}
\label{sec5}

This section explores the diverse applications of Transformer- and \ac{LLM}-based \ac{IDS}. Spanning across computer networks, \ac{IoT}, critical infrastructure, cloud and \ac{SDN}, as well as \ac{AV} and unmanned \ac{UAV}, these environments showcase the adaptive capabilities of modern \ac{IDS} technologies. Highlighting their relevance in safeguarding diverse systems, this section delves into their efficacy and adaptability in various complex and dynamic environments. Besides, Table \ref{tab:Apps} outline the effectiveness and constraints of particular environments and applications in current \ac{SOTA} Transformer and LLM-based IDS implementations.

\subsection{Computer network}

The \ac{IDS} in computer networks are categorized into \ac{HIDS}, which monitors individual devices, and \ac{NIDS}, which analyzes network traffic for threats.
Both examine in different level network logs, which include \textit{tabular features} such as source and destination \ac{IP} addresses, ports, and timestamps, as well as \textit{non-tabular features} like payload data. This payload data often appears in non-human-readable formats such as binary or encoded text. Many research papers have proposed employing
Transformers or \acp{LLM} to address cyber threats in computer network environments. For example, both \cite{steverson2021cyber,unal2022anomalyadapters} analyze log files to address threats. Steverson et al. \cite{steverson2021cyber} leverages DL and NLP on Windows event logs to detect cyber attacks. Data is gathered from an emulated enterprise network undergoing a cyber attack, which involves a spear phishing email and the EternalBlue software exploit to spread botnet malware. A ML anomaly detection algorithm is developed using the Transformer model and self-supervised training. The model demonstrates near-perfect precision and recall in detecting both compromised devices and attack timing. These findings indicate that this method could serve as the detection component of an autonomous endpoint defense system, allowing each device to independently respond to potential intrusions. However, Unal et al. \cite{unal2022anomalyadapters} propose an anomaly adapters scheme, an extensible model for detecting multiple types of anomalies. This model uses the ROBERTa Transformer to encode firewall and host log sequences and employs adapters to learn log structures and identify different anomaly types.

Similarly, the new model, named over-powered (OP) proposed in \cite{duzgun2024network}, combines \ac{MLP} and \ac{CANINE}  architecture to handle numeric, categorical, and text data, respectively. By leveraging \ac{MLP}'s strength in capturing complex relationships in numeric and categorical data and \ac{CANINE}'s character-based encoding for detailed text analysis, the OP model offers can substantially enhance the accuracy of network anomaly detection. Moving on,  the \ac{NIDS} proposed in \cite{zhao2023tbgd}, employed Transformer's \ac{MHA} mechanism and Feedforward neural network to capture global relationships and information. Bi-directional \ac{GRU} is employed to model sequential information in the data, while the \ac{DNN} is designed to learn complex nonlinear relationships, leading to precise intrusion detection predictions. Similarly in \cite{li2022anomaly}, the authors integrate parallel Transformer \ac{GRU} to extract long-distance correlations between timestamps and global features in multivariate time series. This enhances information extraction, thereby improving detection rates for rare anomalies. Alternatively, liu et al. \cite{liu2022collaborative} propose a CIDS-Net architecture to enhance \ac{IDS} performance by integrating features extracted from both network and host log data, encompassing event features and messages. Host data is transformed into vectors using \ac{BERT} word embeddings and aggregated with network features using a fully connected layer. This approach aims to mitigate the challenge posed by the scarcity of datasets containing both network packet and host data. Moreover, \acp{LLM} have been extensively used in \ac{NIDS} tasks across other applications, due to it can learn the characteristics of malicious traffic data \cite{ali2023huntgpt,alkhatib2022can,montes2021web}, capture anomalies in user behaviors \cite{breve2023user}, describe the intent of intrusions and abnormal behaviors \cite{aghaei2023cve,ali2023huntgpt,fayyazi2023uses}, and provide security recommendations and response strategies for identified attack types \cite{chen2023can}.The researchers in  \cite{liu2023malicious} proposed a method to detect malicious \acp{URL} behavior by using \ac{LLM} called  CharBERT to extract hierarchical features of malicious \acp{URL}, extending the application of \acp{LLM} in \ac{IDS} to the user level and demonstrating their generality and effectiveness in intrusion and anomaly detection. CHATAFL scheme \cite{meng2024large}, on the other hand, shifts focus to leveraging \acp{LLM} for generating structured and
sequenced effective test inputs for network protocols lacking machine-readable versions.

Manocchio et al. \cite{manocchio2024flowtransformer} introduced the FlowTransformer framework, a transformative approach for implementing Transformer-based \ac{NIDS}. This framework capitalized on the capabilities of Transformer models like \ac{GPT} 2.0 and \ac{BERT}  to process complex patterns in network traffic, significantly enhancing the detection of sophisticated network threats. \ac{GPT} 2.0, as a generative model, and \ac{BERT}, as an encoder model, were evaluated for their efficacy in this domain, using various network datasets. The principle involved preprocessing network flow data into a format suitable for transformers, analyzing it through configurable transformer components, and efficiently classifying potential threats using a "Last Token" classification head. Notably, the model size could be reduced by over 50\%, and both inference and training times were improved. The best performance achieved was an F1 score of 96.9. Yet, the method's suffer from high dependency on data preprocessing and the challenge of applying appropriate Transformer configurations to network flow data.

\subsection{Internet-of-things}

\Ac{NLP} techniques, encompassing Transformers and \ac{LLM}, have already been employed as a means to build robust and efficient IDS methods in IoT environments. TransIDS \cite{wang2023transids}, a Transformer-based method, adaptively focuses on important features for \ac{IoT} intrusion detection. Moving on, the study \cite{ullah2023tnn} introduces a  \ac{TNN-IDS} for \ac{MQTT}-enabled \ac{IoT} networks. \ac{TNN-IDS} addresses limitations from imbalanced training data by leveraging parallel processing and \ac{MHA} layers in the Transformer, enhancing the detection and learning of malicious activities. In smart home \ac{IoT} scenarios with fewer devices, both network traffic-based and telemetry data-based \ac{NIDS} can work independently. This motivated the researchers to propose a novel ML-based NIDS that combines these approaches in \cite{wang2023securing}. They introduce a Transformer-based \ac{IoT} \ac{NIDS}, which includes self-Attention, to learn attack behaviors from diverse data in heterogeneous \ac{IoT} environments. Other researchers have explored similar techniques for \ac{IIoT}, as documented in references \cite{chai2023ctsf,casajus2023anomaly,yan2023multi,wang2022microcontroller}, akin to previous studies. However, \cite{wang2022microcontroller} utilizes \ac{MCU} temperature fingerprints for \ac{IIoT} intrusion detection. It records temperature sequences, analyzes their relationship with node complexity, computes residuals, and employs a self-encoder model for security assessment. Addressing the escalating challenge of network security, Yan et al. \cite{yan2023multi} propose a multi-Transformer fusion intrusion detection model tailored for real-world \ac{IIoT} environments. {Moving forward, Ferrag et al. \cite{ferrag2024revolutionizing} presented SecurityBERT, a novel BERT-based lightweight architecture tailored for cyber threat detection in \ac{IoT} and \ac{IIoT} networks. Utilizing a privacy-preserving encoding technique and the Byte-Pair Encoding Tokenizer, SecurityBERT efficiently processed network traffic data, including various \ac{IoT} device metrics and network activity logs. The method leveraged \ac{BERT}'s contextual understanding to enhance detection capabilities in \ac{IoT} environments, making it highly suitable for deployment on resource-constrained devices. SecurityBERT achieved impressive performance metrics, setting new benchmarks for real-time traffic analysis in \ac{IoT} devices. For intrusion prediction, Diaf et al. \cite{diaf2024beyond} introduced a novel network intrusion prediction framework designed for \ac{IoT} networks, leveraging the strengths of \ac{GPT}, \ac{BERT}, and \ac{LSTM} models. The principle of the proposed method involved using \ac{GPT} to predict the next network packets based on current ones, \ac{BERT} to evaluate the validity of these predictions, and \ac{LSTM} to classify the packets as normal or malicious. The integration of these models enhanced the framework's predictive capabilities, making it highly effective for IoT cybersecurity.}

\begin{scriptsize}
\begin{longtable}{m{0.3cm}m{0.4cm}m{1.3cm}m{1.2cm}m{4.5cm}m{1.4cm}m{5.2cm}m{0.8cm}}
\caption{Summary of the performance and limitations of specific environments and applications for Transformers-based IDS. In cases where multiple tests are conducted, only the best performance is reported.} 
\label{tab:Apps}\\
\toprule
Ref. & Year & LoTU& Environment & Detected threats & BP (\%)  & Limitations & ADM? \\
\midrule
\endfirsthead

\multicolumn{7}{c}%
{{\bfseries \tablename\ \thetable{} -- Continued from previous page}} \\
\toprule
Ref. & Year & LoTU& Environment & Detected threats & BP (\%)  & Limitations & ADM ? \\
\midrule
\endhead

\midrule \multicolumn{7}{r}{{Continued on next page}} \\ \bottomrule
\endfoot

\bottomrule
\endlastfoot

\hline \toprule

\cite{wang2021ddostc} & 2021 & MHA & Cloud &DDoS, such as reflection and exploitation attacks on \ac{TCP} and \ac{UDP} protocols & AUC= 99.86 & Generalizability unverified due to testing on a single dataset & NA \\\hline

\cite{steverson2021cyber} & 2021 & Transformer  & Computer &  Spear phishing email  & F1= 99.00 & Not tested on a real dataset, thus its generalizability remains unverified. & NA \\\hline

\cite{fu2024iov} & {2024} & {BERT} & {IoV} & {All attacks exists  Car-Hacking, and in-Vehicle IDS datasets, and others. } &  {Acc= 99.97, 99.96} &  {There is a need for more efficient and lightweight BERT models due to limited vehicular resources.} & {NA}\\\hline

\cite{deshpande2023weighted} & 2023 & Transformer  & Computer  & Web attacks via application-layer URLs & Acc= 99.97 & High computational complexity & NA \\\hline

\cite{unal2022anomalyadapters} & 2022 & ROBERTa & Firewall, \newline EC2 nodes & DoS attacks, port scanning, worms, and unidentified machine connections  & F1= 94.5 & Less effective in detecting multiple threats simultaneously & YES \\\hline

\cite{li2022anomaly} & 2022 & TGRU & Server  & Time-series anomalies & F1= 97.81 &Training and testing times were not measured & NA \\\hline

\cite{chai2023ctsf} & 2023 & CNN-Transformer & IIoT & Attacks presented in X-IIoTID dataset  & Acc= 98.87 & Training and testing times were not measured to assess complexity. & UR \\\hline

\cite{casajus2023anomaly}& 2023 & Transformer & IIoT & Attacks presented in WUSTL-IIoT-2021 dataset & F1= 94.31 & Generalizability unverified due to testing on a single dataset & YES \\\hline

\cite{wang2023securing} & 2023 & Self-attention & IoT & Threats are obtained from the ToN IoT dataset. & Acc= 98.39 & The false alarm rate could be further reduced. & NA \\\hline

\cite{wang2022microcontroller} & 2023 & Transformer & IIoT & Temperature & Acc= 89.00 & The detection rate needs improvement & NA \\\hline

\cite{sun2022hierarchical} & 2022 & MHA & Smart grid & Evaluated using attacks present in the NSL-KDD dataset & Acc= 99.48 & Resource heterogeneity can impact the performance of the proposed method & NA \\\hline

\cite{ullah2023tnn} & 2023 & MHA & IoT & Aggressive scan, \ac{UDP} scan, Sparta SSH brute-force, and  MQTT brute-force  & Acc= 99.90 & Generalizability unverified due to testing on a single dataset & NA \\\hline

\cite{wu2022rtids} & 2022 & MHA & Network & Attacks present in CICIDS2017 and CIC-DDoS2019 datasets & F1= 99.17, \newline 98.48 & "Real-time detection has not been verified." & NA \\\hline

\cite{sun2023intrusion} & 2023 & MHA & ICS & Brute Force, DDoS, SQL
injection, XSS  & Acc= 97.24 & Failed to handle zero-day attacks, and generalizability remains unverified & NA \\\hline

\cite{mao2022network} & 2022 & MHA & Smart grid & Evaluated using attacks present in the KDD99 dataset & Acc= 98.03 & Testing solely on one dataset leaves generalizability unverified & NA \\\hline

\cite{diaba2023scada} & 2023 & MHA & SCADA & Bottleneck  & Acc= 99.12 & Lack of generalization to unfamiliar situations. & NA \\\hline
\cite{salam2023deep} & 2023 & MHA & Industry 5.0 & Web-Based Attack such as DoS, SQL injection, and cross-site scripting  & F1= 94.00 & Hybrid methods have not been investigated. & NA \\\hline

\cite{long2024transformer} & 2024 & MHA & Cloud & Botnet, Infilteration, DDoS, DoS, Web, Brute-force & Acc= 93.00 & Generalizability unverified due to testing on a single dataset & UR \\\hline

\cite{alzahrani2022anomaly} & 2022 & MHA & Fog nodes  & All attacks present in UNSW-NB15 dataset & Acc= 98.35 & Data augmentation increases complexity; custom models and extra features may improve accuracy & NA \\\hline

\cite{cobilean2023anomaly} & 2023 & MHA & AV & Malfunction attack & Acc= 99.75 & Generalizability unverified due to testing on a single dataset & NA \\\hline

\cite{duzgun2024network} & 2023 & CANINE & Network & Anomalies in both tabular and text features & MCC= 99.67& Precision remains unimproved; exploring other feature  is recommended. & UR \\\hline

\cite{zhao2023tbgd} & 2023 & Transformer & Network & Attacks in the CICIDS2017 dataset & F1= 100 & Generalizability unverified due to testing on a single dataset. & NA \\ \hline

\cite{liu2022collaborative} & 2022 & BERT and MHA & Network and host & Mitigate the attacks present in the SCVIC-CIDS-2021 dataset & F1= 99.89 & The complexity and detection time are undetermined. & NA \\\hline

\cite{yan2023multi} & 2023 & Multi-Transformers & IIoT & Attacks present in WCIDS  and CICIDS2017 datasets & Acc$\approx$ 98.00   &  WCIDS not publicly available. The aspects of real-time performance and complexity were not measured  & NA\\\hline

\cite{ferrag2024revolutionizing} & {2024} & {BERT} & {IoT, IIoT} & { All attacks present in Edge-IIoTset dataset} & {Acc= 98.2} & {Generalizability is not verified due to training on one dataset.} & {NA} \\\hline

 \cite{diaf2024beyond} & {2024} & {BERT \newline GPT-2} & {IoT} & {All the attacks are present in the CICIoT2023 dataset} & {Acc= 98} & {The model's generalizability remains unverified because it was trained on a single dataset.} & {NA} \\\hline

\cite{fang2022method} & 2022 & Transformer & Network & Attacks in the CICIDS2017 dataset & Acc= 96.10 & Generalizability unverified due to testing on a single dataset. & NA \\\hline

\cite{sun2022informer} & 2022 &  Informer & Network & Assault on an integrated energy system & Acc= 97.80 & The complexity and detection time are undetermined. & NA \\\hline

\cite{hassanin2024pllm} & {2024} & {MHA} & {Satellite} & {Attacks presents in  UNSW-NB-15 and TON\_IoT datasets} & {Acc= 100} & {The method lacks a proper dataset that accurately mimics real satellite system data.} & {NA} \\

\bottomrule
\end{longtable}
\begin{flushleft}
Abbreviations: LLM or Transformer used (LoTU), Inference time (IT); Model size (MS); Deep learning model (DLM); Best performance (BP); Availability of data and materials (ADM), Upon request (UR), not available (NA).
\end{flushleft}
\end{scriptsize}

\subsection{Critical infrastructure}

Critical infrastructure such as smart grids, \ac{SCADA} systems, and web industries are increasingly vulnerable to cyber threats, necessitating robust \ac{IDS}. These systems are crucial for safeguarding against unauthorized access, data breaches, and disruptions that could compromise essential services and operations. The work proposed by Sun et al. \cite{sun2023intrusion} tackles high dimensionality and data imbalance in \ac{ICS} datasets using \ac{IG}-based feature selection and \ac{SMOTE} for oversampling in the aim to increase \ac{IDS} accuracy. The proposed model has been optimized with Bayesian methods, enhancing feature interactions with a \ac{MHA} Transformer and a bi-directional \ac{GRU} to retain temporal features. Moving on, the paper \cite{mao2022network} proposes a network security protection method for power grid information construction, emphasizing multi-service integration. After gathering the information power grid, it quantifies network information risk using attack graphs and analyzes it using a Attention-based Transformer model to detect intrusion types and locations. Finally, it designs a terminal active immune structure using trusted computing to encrypt information and optimize power grid information leakage prevention technology.   Diaba et al. in \cite{diaba2023scada} have introduced an intrusion detection algorithm to tackle this security bottleneck dedicated to \ac{SCADA} systems. The proposed algorithm utilizes the Attention-based Transformer, and the \ac{GSF} feature optimization algorithm, which integrates genetic seeding for enhanced feature selection. The proposed method identifies changes in operational patterns indicative of intruder activity, contrasting significantly with the signature-based approach of traditional \acp{IDS}. 
As Industry 5.0 advances with technologies like AI, IoT, and cyber-physical systems, web-based attack risks rise. Cybersecurity is crucial, as attacks can cause downtime, data breaches, and physical harm. Salam et al. in their research \cite{salam2023deep} propose using \ac{DL} methods, including MHA-based Transformers, \acp{CNN}, and \acp{RNN}, to detect and classify web attacks such as \ac{DDoS}, \ac{SQL} injection, and cross-site scripting that may cause harmful intrusions to the system. Results show that Transformer-based IDS outperforms both \acp{CNN} and \acp{RNN} techniques.

\subsection{Cloud and SDN}

Cloud and \ac{SDN} systems are highly vulnerable to intrusions due to their centralized control, dynamic resource allocation, and multi-tenant environments, which can expose them to attacks like \ac{DDoS}, data breaches, and unauthorized access. Transformers and \acp{LLM} offer significant benefits in enhancing security by efficiently analyzing vast amounts of network traffic data, identifying complex patterns, and detecting anomalies in real-time. Their ability to learn contextual relationships improves accuracy in intrusion detection and mitigates potential threats effectively.  For example, as \ac{SDN} is more susceptible to attacks, particularly severe \ac{DDoS} attacks, it can lead to network collapse. Wang and Li \cite{wang2021ddostc} developed a hybrid neural network, DDosTC, combining Transformers and CNN to detect \ac{DDoS} attacks on \ac{SDN}, and validated it using the CICDDoS2019 dataset. Similarly, Long et al. \cite{long2024transformer} proposed a novel \ac{NIDS} algorithm based on the Transformer model, tailored for cloud environments, enhancing detection accuracy by leveraging the Transformer's Attention mechanism. Moving forward,  Alzahrani et al. \cite{alzahrani2022anomaly} propose an innovative intrusion detection model designed for deployment at fog nodes, aimed at detecting undesirable \ac{IoT} traffic using features from the UNSW-NB15 dataset. Prior to training, correlation-based feature extraction is utilized to lessen computational demand. The Tab Transformer model demonstrates superior performance on continuous data compared to traditional \ac{ML} models and previous benchmarks on the dataset, highlighting its capability with continuous input features.

\subsection{Autonomous vehicles}

\Acp{AV}  face increasing vulnerability to intrusion threats. Addressing these challenges, leveraging \acp{LLM} and Transformers technologies becomes crucial. These advanced models enhance detection and response capabilities, safeguarding these  systemss from evolving cybersecurity risks. Specifically, \acp{AV} infrastructures heavily rely on sensor and electronic component signals, facilitated by wireless technologies that enhance communication but also increase vulnerability to malicious disruptions. The paper \cite{cobilean2023anomaly} proposes a Transformer neural network-based intrusion detection system (CAN-Former IDS) to predict anomalies in \ac{CAN} protocol communications, addressing both the sequence of IDs and message payload values. Advantages include fully self-supervised training and token interaction without hand-crafted features, evaluated using the survival analysis dataset with \ac{CAN} communication from three vehicles. Similarly, Lai study in \cite{lai2023improved}, introduces a federated learning-based edge computing (FL-EC) architecture to enhance privacy and security in \ac{V2X} communications within the \acp{IoV}. By enabling collaborative learning among edge devices without centralizing sensitive data, the FL-EC architecture addresses the limitations of traditional \ac{V2C} systems. The study also presents the feature select Transformer (FSFormer), which employs a feature Attention mechanism to dynamically select significant features for \ac{IDS}, improving the model's ability to extract critical information. Experiments using the UNSW-NB15 and CSE-CIC-IDS2018 datasets showed that FSFormer achieved high accuracy and F1 scores, outperforming other baseline models and \ac{DL} methods. This demonstrates FSFormer's effectiveness in detecting intrusions and securing \ac{V2X} communications.  The paper \cite{fu2024iov} presented IoV-BERT-IDS, a hybrid \ac{NIDS} for the \ac{IoV} in both in-vehicle and extra-vehicle networks, leveraging the \ac{BERT} model. The proposed method involved a semantic extractor that transformed network traffic data, including \ac{CAN} packets, into a format suitable for \ac{BERT} processing. The model then underwent pre-training and fine-tuning to effectively detect intrusions. \ac{BERT} played a critical role in capturing bidirectional contextual features, significantly improving the model's generalization and accuracy in \ac{IoV} environments.

\section{Real-world case studies of Transformers and LLM-based \acp{IDS}}
\label{sec6}

This section examines the reviewed literature, with a particular focus on papers featuring empirical case studies,  \cite{wang2024transformer} for Transformer-based IDS,and \cite{ali2023huntgpt} for LLM-based IDS. Our goal is to assess the application of Transformer and \ac{LLM} algorithms in real-world IDS environments, highlighting their effectiveness and practical deployment, and to offer insights for future replication and extension.

\begin{itemize}
    \item \textbf{Transformer-based method:} In \cite{wang2024transformer}, the authors introduce a similarity-based aggregation algorithm designed to correlate and combine alerts. They then train a Transformer-based model to process variable-length input and complete attack predictions.  The authors constructed a testbed using VMware and vulnerable applications. The network structure is illustrated in Figure \ref{fig:case1}. This testbed includes 4 servers, 2 hosts, essential networking equipment, and various defense devices. To simplify the setup, they configured VMware to use \ac{NAT}  mode, allowing the capture of all traffic between nodes. They simulated a scenario where attackers penetrate an internal network from the outside and perform numerous malicious actions. Different vulnerabilities, such as \ac{XSS} attacks, \ac{SQL} injection, weak passwords, and \ac{RCE}, were embedded in the targeted assets. Additionally, they designed 10 distinct paths to replicate real organized attacks. Given that complex attacks typically involve numerous actions, they simplified the attack paths, focusing on key actions and associated assets. To generate sufficient attack data, two teams black and blue of attackers with different routes were planned, as illustrated in Figure \ref{fig:case1}. The black team initiated the attack, providing data for the training dataset. Subsequently, the blue team  conducted penetration testing, with the resulting data used as the testing dataset. The authors ensured all paths employed similar attack methods, with each attack originating from a different IP address. This approach aimed to verify if the blue team's attacks could be detected after learning from the black team's behavior. A payload sourced from GitHub was used to repeatedly execute blind \ac{XSS}\footnote{\url{https://github.com/payloadbox/xss-payload-list}} attacks and \ac{SQL}\footnote{\url{https://github.com/fuzzdb-project/fuzzdb}} injection attempts.

\begin{figure}[h!]
    \centering
    \includegraphics[scale=0.73]{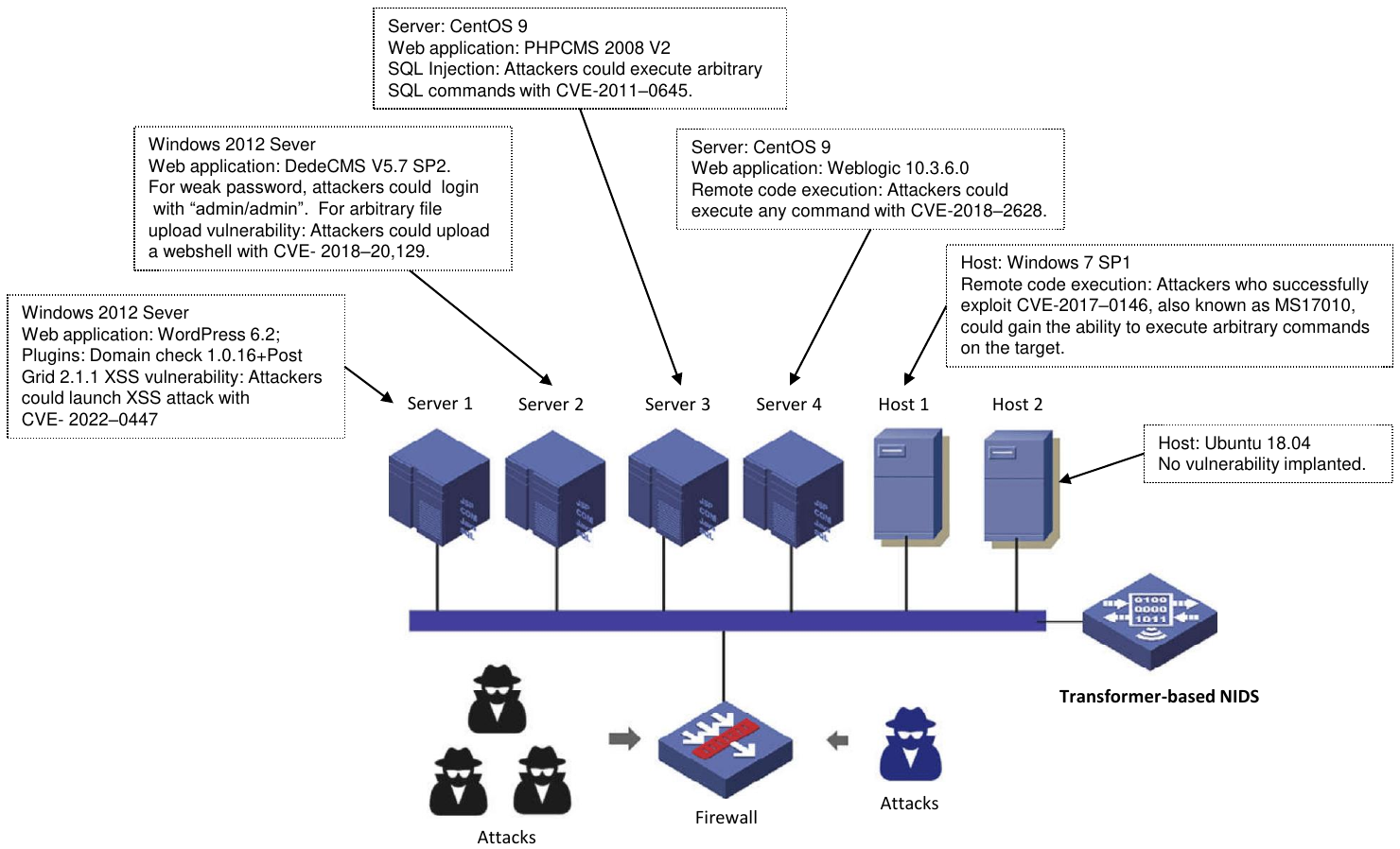}
    \caption{An example of a design of the testbed for deploying Transformer-based \ac{IDS} alert aggregation and attack prediction in a multi-phase attack scenario \cite{wang2024transformer}.}
    \label{fig:case1}
\end{figure}

To generate ample alerts, various security devices, including Firewalls, Transformer-based \ac{NIDS} simulated using Snort, and log analysis tools, such as TCPdump, were used to capture malicious behaviors. Despite efforts to automate attacks, some exploitation required manual intervention. To closely mimic real attackers, extra measures were taken: (i) Attacks were conducted with delays to simulate the entire exploitation process, including discovering security flaws, debugging code, and observing server responses. Random waiting times ensured multiple attack iterations. (ii) No new attack commenced until the previous one was completed, ensuring that attacks did not overlap and alerts were clearly distinguished. (iii) Exploitation scripts were developed based on disclosed reports and open intelligence. For instance, "SQL Injection -> Admin Login," the database name, table name, admin account, and password were sequentially obtained through SQL queries, with all steps incorporated into the exploitation scripts. 

\item \textbf{LLM-based method:} In \cite{ali2023huntgpt}, the authors introduce the development of LLM-b ased IDS scheme, called HuntGPT, a specialized \ac{IDS} dashboard designed to utilize a random forest classifier trained on the KDD99 dataset. The tool incorporates \ac{XAI} frameworks like \ac{SHAP} and \ac{LIME} to enhance the model's usability and interpretability. Combined with a GPT-3.5 turbo conversational agent, HuntGPT aims to present detected threats in a clear and understandable format, emphasizing user comprehension and providing a smooth interactive experience. The system is organized into three distinct layers, each designed to perform specific functions and ensure optimal performance. (i) The \textit{analytics engine} is at the core, responsible for analyzing network packets, identifying anomalies, and processing irregularities within network flows. This layer serves as the powerhouse for network data examination. (ii) The \textit{data storage} layer employs Elasticsearch for its near real-time search capabilities, scalability, and reliability, storing detected anomalies and the corresponding original network data. For storing visual resources like plots and images, Amazon S3 buckets are used, providing security and easy access. (iii) The \textit{user interface}, built with Gradio, functions as the interactive front-end of the system. It presents the outcomes from the analytics engine in a user-friendly manner and integrates with OpenAI’s \ac{LLM} API, facilitating seamless interactions between analysts and the system for ongoing discussions and analysis. Figure \ref{fig:case2} illustrates the hardware and software required for building an \ac{LLM}-based IDS suggested in \cite{ali2023huntgpt}.

\begin{figure}[h!]
    \centering
    \includegraphics[width=0.7\linewidth]{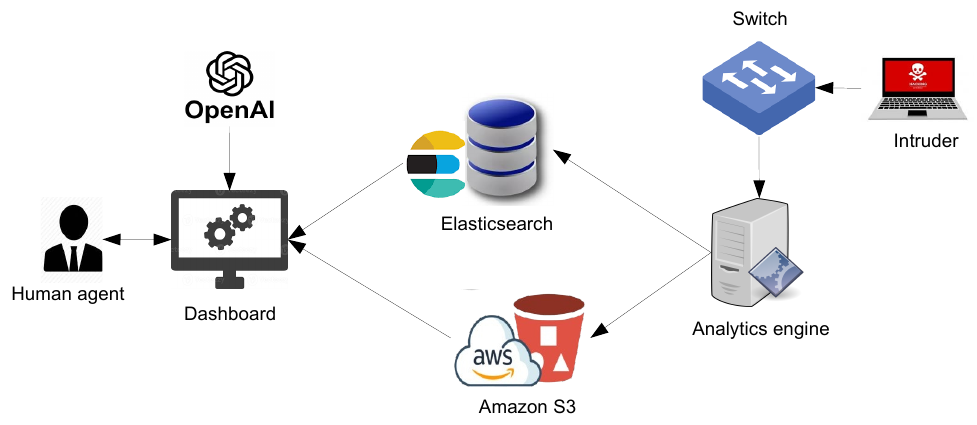}
    \caption{Hardware interconnection, detailing the functions and interactions of each component within the overall system architecture. }
    \label{fig:case2}
\end{figure}

The system's modular design allows each layer to be developed, maintained, and scaled independently, ensuring flexibility and efficient scalability. The autonomous nature of these components contributes to the system’s robustness and adaptability.

The \textbf{\textit{anomaly detection application server}} serves as the central orchestrator of the anomaly detection process. It integrates several sub-modules, starting with the \textit{ML model loader}, which loads a pre-trained machine learning model trained on the KDD99 dataset to detect anomalies and provide explainable insights using \ac{SHAP} and \ac{LIME}. The \textit{Elasticsearch connector} handles secure communication with Elasticsearch for data storage and indexing, while the \textit{prediction} component analyzes network flows to identify anomalies. The \textit{explainer} generates explanations, plots, and JSON documents, indexing data into Elasticsearch and uploading plots to AWS S3 to enhance understanding of the model's behavior. The \textbf{\textit{\ac{IDS} dashboard}} enhances user trust by providing detailed model insights and explanations. It allows users to inspect original data packets for manual anomaly detection and facilitates interactive discussions with the AI assistant. This component integrates several sub-modules, including the \textit{OpenAI connector} for authentication and conversation tracking, the \textit{anomaly packet data fetching} module for extracting relevant information from Elasticsearch, the \textit{OpenAI API unit} for integrating data with prompts for OpenAI, and the \textit{AI assistant analysis} module, which generates comprehensive analyses and facilitates interactive communication with human analysts.
\end{itemize}

Deploying Transformer and \ac{LLM}-based \ac{IDS} in real-world scenarios presents a set of obstacles that can significantly affect their efficacy and feasibility. For instance, the complexity and computational demands of these models, as exemplified in the case studies using VMware setups and simulated attacks, indicate substantial hardware requirements. Such \ac{IDS} implementations often necessitate robust servers and a substantial computer network and information technology infrastructure, which may not be viable for all organizations, especially smaller ones with limited budgets. Additionally, the real-time processing capabilities of these systems are crucial for effective network security but can be hampered by the hardware-intensive nature of Transformer and \acp{LLM} algorithms. For example, processing vast amounts of data in real-time requires high-performance computing environments that can handle the intensive workloads generated by these sophisticated models. This requirement may lead to significant energy consumption and increased operational costs, which could be prohibitive.

Another critical limitation is the adaptability of these systems to different network environments. The testbed configurations, which often include a variety of network scenarios and attack vectors, must be meticulously designed to mirror the diverse real-world network landscapes. However, this can be challenging as each network environment has unique characteristics and threat profiles that may not be fully addressed by a standardized \ac{IDS} solution. Moreover, maintaining the accuracy of such systems in dynamic environments where new threats continuously emerge is problematic. The models must be constantly updated and trained on the latest threat data to remain effective, which involves continuous data collection and analysis, further straining resources.

\section{Research challenges and future directions}
\label{sec7}

The application of Transformers, specifically \acp{LLM}, in cybersecurity represents a cutting-edge frontier, showcasing their robust capabilities in addressing complex and dynamic cyber threats. However, despite their strengths, these models face significant challenges. The following analysis aims to illuminate the complexities of leveraging \acp{LLM} in cybersecurity by analyzing these highlighted challenges. It underscores the need for further investigation and broader application of \acp{LLM} in future cybersecurity efforts.

\subsection{Transformers challenges}

Applying Transformer-based models for \ac{IDS} across different environments such as computer networks, \ac{IoT}, industrial infrastructures, \ac{SDN}, and host logs presents a unique set of challenges: \\

\noindent \textbf{(a) Variations in intrusion categories:} Transformers-based \ac{IDS} face challenges in effectively addressing intrusions and threats in both signature-based and anomaly-based scenarios. In signature-based detection, where known patterns are matched against incoming data, Transformers must efficiently process and compare large volumes of data to detect malicious signatures in real-time, requiring substantial computational resources. Meanwhile, anomaly-based detection, which identifies deviations from normal behavior, demands robust understanding of contextual nuances to distinguish between genuine anomalies and false positives. Transformers' ability to capture intricate contextual information can enhance anomaly detection but necessitates continuous training and adaptation to evolving threats. 

\noindent \textbf{(b) Data heterogeneity}:  Computer networks generate data like \acp{PCAP}, flow records, and log files, each with different structures and patterns. In \ac{IoT} environments, devices produce data with varying formats, volumes, and transmission frequencies, and low-power devices might have limited logging capabilities. Industrial infrastructures often rely on specialized protocols and legacy systems, necessitating specialized knowledge for data preprocessing. \acp{SDN} generate data from control plane messages, flow statistics, and configuration changes, which differ significantly from traditional network data. Host logs vary widely in format and content, from application logs to system event logs. Transformers typically require consistent data formats and structures for effective learning and inference.

\noindent\textbf{(c) Scalability and real-time processing:} As Transformers, particularly large models, require significant computational resources. Real-time or near-real-time processing of large-scale data in environments like \ac{SDN} or industrial infrastructures can be challenging. Additionally, while Transformers excel at learning patterns and detecting anomalies, many environments still depend on signature-based detection, and effectively combining these approaches can be complex. Real-time processing demands further complicate the use of Transformers. Environments like \ac{SDN} and \ac{IoT} often require real-time or \ac{URLLC} \cite{kheddar2022efficient} to  responses to intrusions, which can be difficult to achieve due to the computational demands of Transformer models. Ensuring low latency and high throughput is critical in environments like \ac{SDN} and computer networks, where delay-sensitive applications are common. However, Transformers are often considered "black boxes," making it difficult to explain their decisions. This lack of interpretability can be a significant drawback in environments where understanding the cause of an alert is crucial.

\noindent \textbf{(d) Label scarcity and data privacy:}  Label scarcity and imbalanced datasets present additional challenges. Obtaining labeled datasets for training can be difficult, particularly for rare intrusion events. Many environments suffer from highly imbalanced datasets where malicious activities are rare compared to normal operations. Moreover, Transformer models can be susceptible to adversarial attacks, where attackers craft inputs to deceive the model, leading to false negatives or false positives. Data privacy, especially in industrial and \ac{IoT} environments, where handling sensitive data requires stringent privacy and security measures. Ensuring compliance with regulations like \ac{GDPR} adds another layer of complexity. Integrating Transformer-based models with existing security infrastructure, such as security information and event management systems and firewalls, can be complex and may require significant changes to existing workflows. Effective application of Transformer models also requires domain-specific knowledge for feature engineering, data preprocessing, and interpreting results, which is particularly challenging in specialized environments like industrial infrastructures.

\noindent \textbf{(e) Updating Transformers: } Is crucial to adapt to evolving threats and environmental changes. However, this process can be resource-intensive and challenging to manage. In resource-constrained environments like \ac{IoT}, deploying computationally heavy Transformer models may be impractical. Thus, optimizations or adaptations are necessary to fit the model within limited hardware capabilities. Balancing model efficacy with resource constraints is essential for ensuring efficient and effective use of Transformer-based solutions in such contexts.

\subsection{LLMs challenges}

Since \acp{LLM} are built on Transformer-based architecture, they can be viewed as an enhanced version of Transformers. Therefore, all challenges associated with Transformers also valid to \acp{LLM}. Additionally, there are specific challenges unique to \acp{LLM}, as outlined below: 

\noindent\textbf{(a) Data challenges and privacy concerns:} The quality, diversity, and volume of data significantly affect the performance and generalization capabilities of these models. Due to their scale, \acp{LLM} typically require large amounts of data to capture nuanced distinctions, but obtaining such data can be challenging. Many specific security tasks lack high-quality and robust publicly available datasets. Using limited or biased datasets may cause models to inherit these biases, resulting in skewed or inaccurate predictions. Additionally, there is a risk of benchmark data contamination, where redundant filtering of native data in existing research could lead to overlap between training and testing datasets, inflating performance metrics. Moreover, the researchers have serious concerns about the inclusion of personal information, such as phone numbers and email addresses, in training data for \acp{LLM} used in information and content security tasks, which could lead to privacy breaches during the prompting process.

\noindent\textbf{(b) Attacks targeting LLMs:} \acp{LLM} face significant vulnerabilities, categorized into backdoor and prompt injection attacks. Backdoor attacks like ICLAttack \cite{zhao2024universal} and BadGPT \cite{shi2023badgpt} manipulate \ac{LLM} outputs by embedding triggers into the model or its inputs, enabling malicious behavior without direct fine-tuning. Prompt injection attacks, such as \ac{P2SQL} \cite{pedro2023prompt} and \ac{CIA} \cite{jiang2023prompt}, exploit \acp{LLM}  by inserting malicious commands disguised as benign prompts, compromising data integrity and generating harmful content. Techniques like HOUYI \cite{sun2024llm4vuln} automate these attacks across diverse scenarios, highlighting the need for robust defenses to secure \acp{LLM} from manipulations. \acp{LLM} are also susceptible to jailbreaking attacks, where malicious prompts induce them to produce harmful outputs despite security measures. Efforts to mitigate these risks, including genetic algorithms and semantic firewalls, struggle against evolving attack sophistication \cite{zhang2024llms}. Beyond direct attacks, generative AI and \acp{LLM} introduce ethical and cybersecurity challenges, ranging from deceptive behaviors under specific triggers to vulnerabilities in application integrity and data privacy. These findings underscore the urgent need for ongoing research and proactive security measures to safeguard \acp{LLM} against evolving threats, ensuring their responsible deployment in cybersecurity and beyond.

\noindent\textbf{(c) Prompt engineering for LLM-based IDS:} Applying \acp{LLM} to \ac{IDS}, prompt engineering poses several challenges. Firstly, \acp{LLM} like GPT-3 require massive computational resources, which can be prohibitive for real-time \ac{IDS} applications. Fine-tuning these models for security-specific tasks necessitates large, labeled datasets, often scarce in cybersecurity. \acp{LLM} may struggle with domain-specific jargon and nuanced threat contexts, impacting detection accuracy. Interpretability remains a concern, as \ac{LLM} decision-making processes can be opaque. Adapting prompts to elicit relevant security insights without compromising model performance requires expert knowledge. Lastly, ensuring \acp{LLM} comprehend adversarial tactics and remain resilient to evasion techniques is crucial for robust \ac{IDS} deployment.

\noindent\textbf{(d) Generalization capability of \acp{LLM}:} The generalization capability of \acp{LLM} pertains to their ability to consistently and accurately execute tasks across diverse datasets or domains beyond their training environment. Despite undergoing extensive pre-training on large datasets to acquire broad knowledge, the absence of specialized expertise can present challenges when \acp{LLM} encounter tasks beyond their pre-training scope, especially in the IDS domain. As discussed in \cite{xu2024large}, the authors have explored the utilization of \acp{LLM} in 21 security tasks spanning five security domains. They observed substantial variations in the context and semantics of code or documents across different domains and task specifications. To ensure \acp{LLM} demonstrate robust generalization, meticulous fine-tuning, validation, and continuous feedback loops on datasets from various security tasks are imperative. Without these measures, there’s a risk of models overfitting to their training data, thus limiting their efficacy in diverse real-world scenarios.

\noindent\textbf{(e) Trust, interpretability, and ethical application of \acp{LLM}:} Establishing trust in \acp{LLM} for \ac{IDS} requires developing technologies and tools that provide deeper insights into model internals, enabling developers to understand the reasoning behind generated outputs. Improving interpretability and trustworthiness can facilitate the widespread adoption of cost-effective automation in cybersecurity, thereby enhancing the efficiency and effectiveness of security practices in \ac{IDS} applications. Many \acp{LLM} used in \ac{IDS} lack open-source availability, raising concerns about the quality, sources, and ownership of their training data, which in turn raises questions about task ownership and data integrity. Furthermore, the susceptibility to adversarial attacks poses a significant threat, as techniques to manipulate \acp{LLM} can potentially compromise security measures and expose sensitive training data.

Ensuring interpretability and ethical implementation is critical when integrating \acp{LLM} into \ac{IDS} tasks, given their opaque nature and the sensitivity of security requirements. Understanding how these models make decisions is challenging, impeding explanations for generated outputs and recommendations in \ac{IDS} contexts. Concerns must be highlighted and addressed to effectively mitigate additional security risks associated with artificial intelligence-generated content in \ac{IDS} including fake content \cite{yu2024fake},  privacy breaches, dissemination of misinformation, and the creation of exploitable code. The lack of interpretability and trustworthiness may lead to user uncertainty and skepticism, as stakeholders may hesitate to rely on \acp{LLM} for \ac{IDS} tasks without clear insights into their decision-making processes or adherence to rigorous security standards.

\noindent\textbf{(f) Transfer learning from pre-trained \ac{LLM}: } Pre-trained models such as GPT and BERT reveal a range of limitations and potential biases that are critical to consider. One of the primary limitations is the heavy reliance on extensive and diverse datasets for training these models. While pre-trained models have demonstrated remarkable capabilities in understanding and generating text based on their training, they are often constrained by the data they were trained on. This dependency can lead to biases in model outputs, particularly if the training data was not balanced or if it contained inherent prejudices. For instance, models like \ac{GPT} and \ac{BERT} may inadvertently perpetuate or amplify biases present in their training sets, leading to skewed or unfair outcomes when deployed in real-world applications. Moreover, these models require significant computational resources not only for training but also for fine-tuning and deployment, which can be a barrier for institutions with limited hardware capabilities. This creates a divide in who can effectively use these technologies, potentially widening the gap between large organizations and smaller entities.

Another concern is the adaptability of these models to specific tasks or domains. While they are generally powerful, transferring a pre-trained model to a specialized domain without extensive retraining can lead to suboptimal performance. This limitation necessitates additional layers of fine-tuning, which again demands more data and computational power, further complicating their use in varied applications. These challenges highlight the need for ongoing research to mitigate biases, reduce resource demands, and enhance the flexibility of pre-trained models for broader and more equitable usage.

\subsection{Perspectives and future directions}

Since \acp{LLM} represent the latest advancement in Transformer technology, future research and perspectives mostly focus exclusively on \acp{LLM}, {but are also applicable to Transformers}. Despite extensive research into \acp{LLM} within cybersecurity, their exploration and application are still in the early stages, offering considerable potential for growth \cite{motlagh2024large}. The complexity of cybersecurity arises not only from the variety of attack methods but also from the intricate nature of network environments, combined with the necessity for a comprehensive application of diverse tools and strategies to ensure effective protection \cite{azizi2023cybersecurity,zhang2024llms}. Addressing these challenges requires AI systems with advanced capabilities in planning, reasoning, tool use, memory, and more. Building upon this discussion, the authors identify several critical perspectives that need to be addressed in LLM-based \ac{IDS}, as follows:

\noindent\textbf{(a) Enhanced efficiency and explainable decision-making:} 
{Transformer-XL \cite{dai2019transformer}, known for its ability to capture long-term dependencies through segment-level recurrence, can be further optimized for IDS by enhancing its understanding of temporal contexts. \ac{IDS} requires analyzing sequences of network events over time to detect anomalies effectively. Future research could focus on adapting Transformer-XL to better handle time-series data typical in network traffic. This could involve developing novel mechanisms for capturing time-dependent patterns in network flows, improving the model's accuracy in identifying slow and stealthy attacks that unfold over extended periods. In addition, Longformer \cite{beltagy2020longformer} introduces the concept of sparse attention, which reduces computational complexity by focusing only on relevant parts of the input sequence. This feature is particularly beneficial for IDS, where real-time analysis of massive volumes of data is essential. Future work could explore optimizing Longformer’s sparse attention mechanism to handle diverse types of network traffic efficiently. Researchers could also investigate hybrid models that combine the strengths of Transformer-XL’s recurrence with Longformer’s sparse attention, creating a more scalable and efficient IDS model capable of real-time intrusion detection in large-scale networks.} Moreover, 
future \acp{LLM} for \ac{IDS} need to focus on advancing contextual understanding of security events and network behaviors. This includes training models to interpret complex relationships between network activities, identify anomalous patterns indicative of potential threats, and differentiate between benign and malicious activities with greater accuracy. Simultaneously, there will be efforts to enhance interpretability and explainability of \ac{LLM} decisions within \ac{IDS} frameworks. This involves developing techniques to make \ac{LLM} outputs more transparent, enabling security analysts to understand how decisions are made and fostering trust in automated \ac{IDS} processes. By improving both contextual understanding and explainability, future \acp{LLM} can elevate the effectiveness and acceptance of \ac{IDS} in detecting and responding to cybersecurity threats.

\noindent\textbf{(b) \ac{RL} with \ac{LLM} agent:} \ac{RL} is a \ac{ML} paradigm where an agent learns to make decisions by interacting with an environment to maximize cumulative rewards. A Q-table is a matrix contains Q-values, that represents the expected future rewards for actions taken in particular states. An agent is the decision-maker in \ac{RL}. A Q-network, on the other hand, is a neural network that approximates the Q-values, allowing \ac{RL} to scale to larger state-action spaces \cite{gueriani2023deep}. Applying \ac{LLM} agents in \ac{IDS} can enhance adaptability and understanding of complex patterns. However, \ac{LLM} agents do not replace classical \ac{RL} agents but can complement them by providing richer contextual understanding. \acp{LLM} do not replace Q-tables directly; instead, they can augment \ac{RL} by informing better state and action representations, leading to more effective learning and decision-making. The emergence of LLM-based agents presents a compelling perspective in the realm of \ac{IDS}. An LLM-based \ac{IDS} agent is conceptualized as a system that utilizes a \ac{LLM} to analyze network behaviors, detect intrusion patterns, and respond effectively using a variety of tools \cite{pankajakshan2024mapping}. By harnessing advanced \ac{NLP} capabilities, \ac{LLM} agents introduce novel approaches to cybersecurity \cite{yan2024depending,zhan2024injecagent,xi2023rise}. They hold the potential to significantly enhance the efficiency of threat detection, defense strategy formulation, and adaptation of security policies within IDS environments. \ac{LLM} agents, equipped with frameworks that integrate perception, action, and real-world interactions through APIs and tools \cite{xi2023rise}, offer promising avenues for automating detection tasks, improving response times, and managing complex security incidents. However, their deployment in \ac{IDS} contexts necessitates addressing inherent security risks, including novel threats such as Web-based indirect prompt injection \cite{zhan2024injecagent}. Continued research into \ac{LLM}-based \ac{IDS} agents is crucial for advancing adaptive, intelligent, and robust cybersecurity defenses.

\noindent\textbf{(c) Enhancement of \acp{LLM} for IDS: } The evolution of \ac{IDS} research involves critical decisions between leveraging pre-trained models like GPT-4 and open-source frameworks such as T5 or LLaMa. GPT-4 offers rapid customization for \ac{IDS} tasks with minimal data, reducing computational overhead but limiting extensive retraining capabilities. Conversely, frameworks like T5 provide extensive customization through retraining on large datasets, demanding significant resources yet enabling the development of highly specialized \ac{IDS} models. Enhancing \ac{IDS} effectiveness through inter-model collaboration entails integrating multiple \acp{LLM} or combining them with specialized \ac{ML} models to streamline complex security tasks and boost efficiency. Notably, ChatGPT has emerged as a valuable tool in \ac{IDS} research \cite{daniel2023labeling,xie2023defending}, due to its computational efficiency, versatility, and potential cost-effectiveness compared to other \acp{LLM}. These advancements underscore the transformative potential of \acp{LLM} in shaping the future of \ac{IDS}, paving the way for more collaborative, efficient, and adaptive cybersecurity solutions.

\noindent\textbf{(d) Multimodal inputs of \acp{LLM} for IDS:}
In the previous section, it was observed that \ac{ViT} utilized 2D inputs such as images and matrices. Additionally, novel CNN-Transformers approaches have been proposed to convert \ac{IDS} datasets into 2D representations for training efficient models capable of detecting attacks. Moreover, GAN-Transformers are employed with imbalanced 2D data to generate balanced attack datasets. In security applications, \acp{LLM} typically utilize input from code-based and text-based datasets. The introduction of new input formats rooted in these \ac{NLP} algorithms, such as image inputs, alongside text, presents an exciting opportunity to enhance \acp{LLM}' capability in detecting intrusion threats. Images can effectively illustrate security processes and requirements, offering \acp{LLM} additional perspectives. Furthermore, multimodal inputs combining text and visuals provide a more comprehensive contextual understanding, resulting in more precise and contextually relevant security solutions. This expansion into underdeveloped domains holds significant potential for advancing automated security solutions. {For example, future advancements could involve integrating Transformer-XL and Longformer with multimodal learning techniques to process and correlate information from multiple sources. This integration would provide a more comprehensive view of network activity, enhancing the models’ ability to detect complex attack patterns that might be missed when analyzing data in isolation. Researchers could explore architectures that allow these models to simultaneously process textual logs, numerical features, and graphical representations of network traffic.}

\noindent\textbf{(e) Enhancing IDS with \acp{LLM} and data hiding analysis:} Data hiding, specifically steganography and covert channels, are alternative techniques to cryptography, enabling senders and receivers to exchange secret data in unconventional ways. The work in \cite{kheddar2019pitch,kheddar2022high} classifies these techniques into two main categories: protocol packet modification, which involves altering the payload \cite{kheddar2022speech}, protocol-specific fields \cite{chen2023covert}, or both; and modification of packet time relations which includes changing: packet transmission order, inter-packet delay, or packet drop rate. These techniques inherently pose a risk of causing intrusions in computer networks \cite{li2023detecting}, particularly when countermeasures like covert channel detection \cite{li2023detecting,caviglione2024you} are selectively applied across network entities. \acp{LLM} can assist \ac{IDS} in analyzing packet flow behaviors and detecting fields susceptible to covert channel attacks, given that the captured information typically manifests as either text or tabular data.  For instance, \acp{LLM} can trigger alarms when \ac{TTL} parameters exceed a specific threshold, indicating unusual delays or potential risk of packet drops. Moreover, \acp{LLM} can scrutinize IP packets for sequence number alterations; irregular increments could prompt \acp{LLM} to raise alarms. Additionally, \acp{LLM} can monitor for anomalies in packet header checksums, unexpected changes in packet payload sizes, or deviations in packet transmission times, all of which may signify covert channel activity.

\noindent\textbf{(f) Selective TL among \acp{LLM} for IDS:} \ac{TL} leverages pre-trained models, adapting their knowledge to new but related problems, thus saving time and computational resources. This approach is especially useful in \ac{DL} applications with limited and scarce data \cite{himeur2023video}. {Transformer-XL and Longformer could be adapted to leverage unsupervised learning techniques, allowing them to identify patterns and anomalies in unlabeled data. Future research could focus on developing self-supervised pre-training tasks specific to network security, such as predicting network event sequences or reconstructing corrupted traffic logs. These pre-training tasks would enable the models to learn robust representations of normal and malicious network behavior, improving their detection capabilities even in the absence of labeled data.} In addition, leveraging knowledge from \acp{LLM} specifically designed for domains closely related to \ac{IDS}, such  TelecomGPT \cite{zou2024telecomgpt}, can significantly enhance IDS performance. Domain-specific \acp{LLM}, like GenAI models \cite{bariah2024large}, excel in telecommunication tasks such as network optimization, sensing, protocol analysis, transmission, fault detection, and more. By transferring this expertise to an \ac{LLM} dedicated to \ac{IDS}, researchers can capitalize on the overlapping concepts of networking, protocols, and security. This \ac{TL} approach results in a more robust \ac{IDS} compared to utilizing knowledge from general \ac{NLP} \acp{LLM} like GPT-4, as it provides a deeper understanding of network behaviors and anomalies. For instance, insights from telecom \acp{LLM} about network traffic patterns can directly inform \ac{IDS} detection algorithms, enhancing their ability to distinguish malicious activities from normal operations with greater precision.

\noindent\textbf{(g) Integration with dynamic security Ecosystems and adversarial resilience:} Future \acp{LLM} for \ac{IDS} need to increasingly integrate with dynamic security ecosystems, incorporating real-time threat intelligence from diverse sources such as threat feeds, vulnerability databases, and incident reports. {For example, the suggested EvolveDroid system \cite{huang2024strengthening} enhances \ac{LLM} ecosystem security by leveraging multiple feature types and contrastive learning to enhance detection capabilities and adapt to evolving threats.}  This integration not only enriches model training but also enhances adversarial resilience by enabling \acp{LLM} to detect and mitigate sophisticated adversarial attacks. Additionally, collaborative defense systems where multiple specialized \acp{LLM} work synergistically will further strengthen \ac{IDS} capabilities, improving overall threat detection accuracy and reducing false positives. This holistic approach ensures that LLM-based \ac{IDS} are well-equipped to handle evolving cybersecurity challenges effectively. {Nevertheless, the field is still in the early stages of research investigation and requires further exploration and development to fully understand its implications and potential.}

\noindent\textbf{(h) Cross-domain applications of Transformer and LLM-based IDS:} {As highlighted by the limited cross-domain efficacy of traditional \ac{DL}-based \ac{NIDS} models \cite{layeghy2023explainable,houssel2024towards}, there is a need for more resilient and flexible architectures. The future direction of utilizing Transformers and \acp{LLM} in \ac{IDS} shows promise for cross-domain applications, especially in sectors like healthcare and finance, which face significant network security challenges \cite{chen2024survey2}.
In healthcare, fine-tuned \acp{LLM} \cite{cui2024anytasktune} can be specifically tailored to detect and analyze anomalies and potential threats in sensitive medical data, ensuring patient data integrity and confidentiality. This is critical as the sector handles vast amounts of private information that, if compromised, could have severe consequences. Utilizing Task-Fine-Tune allows these models to precisely identify and react to unique patterns of cyber threats that are specific to the healthcare domain, such as ransomware attacks on hospital systems or unauthorized access to medical records. Similarly in finance \cite{chen2024survey2}, \acp{LLM} optimized for IDS tasks can monitor and analyze transactional data for signs of fraudulent activities with heightened accuracy. Financial institutions are often targets of sophisticated cyber-attacks aimed at financial gain. By fine-tuning \acp{LLM} on specific financial datasets, such as those involving transactional behaviors and fraud indicators, these models can become highly effective at detecting and mitigating potential threats, thereby safeguarding assets and sensitive financial information. 
Both sectors benefit from the ability of \acp{LLM} and Transformers to process and analyze large volumes of text and other data types quickly and accurately. This capability allows for real-time threat detection and enhances the overall security infrastructure of organizations in these critical fields. By incorporating these advanced technologies, healthcare and finance institutions can better protect sensitive data and ensure compliance with regulatory standards, ultimately safeguarding both individual privacy and financial integrity. However, the use of \ac{LLM}-based \ac{IDS} in healthcare and finance remains an under-researched area that warrants further exploration in future studies.}

\noindent\textbf{(i) Interpretability,  scalability, {and real-time improvement}:}
Integrating domain knowledge into Transformer and \ac{LLM}-based \ac{IDS} offers a promising direction for future research. By incorporating specific insights and contextual information from various fields, the interpretability of these models can be significantly improved. This approach allows for a deeper understanding of the patterns and anomalies detected, making the results more meaningful and actionable for cybersecurity experts. For instance, in the healthcare sector, integrating medical domain knowledge can help the model distinguish between normal and abnormal network activities related to medical devices and patient data flows. Similarly, in the financial sector, understanding the typical transaction patterns and identifying deviations can enhance the model's ability to detect fraudulent activities.

Additionally, employing knowledge distillation \cite{gou2021knowledge} techniques can further enhance the efficiency and scalability of these \ac{IDS} models. Knowledge distillation involves training a smaller, more efficient model (student) to replicate the performance of a larger, more complex model (teacher). This process can significantly reduce the computational resources required while maintaining high detection accuracy. Furthermore, designing a more efficient and scalable model architecture is crucial. Future research should focus on optimizing Transformer and \ac{LLM} architectures to handle large-scale data more effectively. Techniques such as sparse attention mechanisms, model pruning \cite{zhu2017prune}, and distributed computing can be explored to reduce computational overhead and improve real-time processing capabilities. Combining domain-specific knowledge with advanced \ac{IDS} architectures and knowledge distillation not only enhances the model's interpretability but also ensures that the system remains robust and scalable. This holistic approach can lead to more reliable and efficient \ac{IDS}, capable of addressing the unique security challenges across different industries.

{Moreover, the development of pre-trained Transformers and \ac{LLM}s tailored for \ac{IDS} facilitates the creation of lightweight models adept at real-time threat processing. For example, \acp{SLM} such as TinyLlama \cite{zhang2024tinyllama} offer compact versions of \acp{LLM}, engineered to provide similar capabilities for specialized tasks with significantly reduced computational and memory demands. These models proficiently process and interpret both text and tabular data. In real-time intrusion processing, \acp{SLM} could be used to rapidly analyze vast quantities of log data, enabling quicker and more effective threat detection with minimal hardware investment. This expedited processing capability is crucial for immediate threat responses in cybersecurity environments. Researchers are urged to explore the potential of real-time Transformers and \acp{LLM}-based \ac{IDS} by applying their methods on embedded platforms such as \acp{FPGA}, and by measuring real-time metrics including inference time, latency, resource utilization, and algorithmic complexity. Such evaluations provide valuable insights for users and developers, helping determine the most suitable Transformer or \ac{LLM} for specific hardware setups and IDS threat types. }

\section{Conclusion}
\label{sec8}

The field of intrusion detection is rapidly evolving, driven by advancements in Transformer-based models. These models, with their powerful Attention mechanisms, have outperformed traditional detection systems, offering enhanced capabilities in identifying subtle anomalies within complex network environments. However, several challenges still limit their full potential, including issues related to data diversity, real-time processing, computational resources, and model interpretability. The survey underscores the growing importance of integrating Transformers into \acp{IDS} to improve cybersecurity, particularly by capturing both spatial and temporal anomalies through models incorporating \ac{CNN}/\ac{LSTM} layers. These approaches reduce false positive rates and enhance detection accuracy. This survey also explores novel Transformer-based methods, such as \ac{ViT}-based and \ac{GAN}-Transformer-based \acp{IDS}, which offer advanced capabilities in detecting complex attack patterns and improving model performance through synthetic data generation, showcasing the versatility of Transformers in strengthening \acp{IDS}.

Moreover, the relevance of Transformers extends beyond conventional \acp{IDS}. Transformer-based models, such as those employing \acp{LLM} for intrusion detection, show significant potential in addressing challenges associated with heterogeneous data sources,  and advanced attack vectors. By utilizing context-aware models like GPT-based or BERT-based \acp{IDS}, these systems offer improved adaptability in complex cybersecurity environments. More over, this survey includes a dedicated section on real-world case studies of Transformers and \acp{LLM} in \acp{IDS}. This section examines the reviewed literature, focusing on empirical case studies for Transformer-based \acp{IDS} and \ac{LLM}-based \acp{IDS}. By assessing the application of these models in real-world \ac{IDS} environments, the section highlights the practical effectiveness of these algorithms, their deployment challenges, and provides valuable insights for future replication and extension in operational settings.

Despite the challenges outlined in this survey, both Transformers and \acp{LLM} offer immense potential for advancing the field of intrusion detection. As research in this domain continues to evolve, it is anticipated that these models will play a critical role in enhancing the security, privacy, and resilience of digital infrastructures. Future innovations in Transformer and \ac{LLM}-based \acp{IDS}, alongside advancements in data handling, real-time processing, and model interpretability, promise to revolutionize intrusion detection, ensuring these systems can effectively tackle the ever-evolving cybersecurity threats.

\printcredits

\section*{Declaration of competing interest}
The author declare that they have no known competing financial interests or personal relationships that could have appeared to influence the work reported in this paper.

\section*{Data availability}
No data was used for the research described in the article.

\section*{Acknowledgement}

The author acknowledges that the study was partially funded by the Algerian Ministry of Higher Education and Scientific Research (Grant No. PRFU--A25N01UN260120230001).

\bibliographystyle{elsarticle-num}
\bibliography{references}

\newpage

\bio{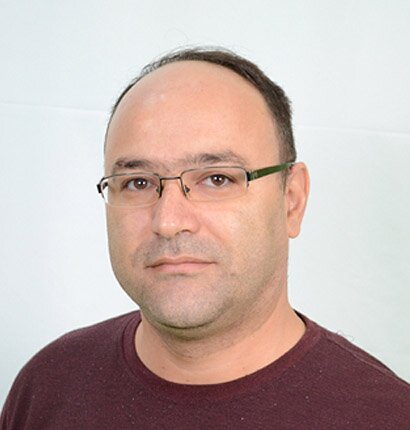}
\textbf{Dr. Hamza Kheddar, SMIEEE} is an Associate Professor at the University of Medea and a Researcher at the LSEA Lab, Medea, Algeria. He holds a Ph.D. in Telecommunication from USTHB University. Dr. Kheddar is an expert in speech steganography, digital watermarking, intrusion detection, biometrics, and more. In June 2022, he obtained the Habilitation to Direct Research. He serves as a reviewer for esteemed journals including Computers \& Security, Computer Speech \& Language, Applied Intelligence, Knowledge-Based Systems, Information Fusion, Speech Communication, IEEE Access, IEEE TIFS, and IEEE Communications Letters. Dr. Kheddar has authored more than 25 papers and contributes significantly to diverse research areas such as image classification, intrusion detection, artificial intelligence, transfer learning, reinforcement learning, and 5G/6G slicing and security. Additionally, he serves as a telecommunication track chair at several conferences, including IC2M 2023 and ICTSS 2024.
\endbio

\end{document}